\documentclass[preprint,12pt,1p]{elsarticle}
\usepackage{subfigure}
\usepackage{amsmath}
\usepackage{amssymb}
\usepackage{url}
\biboptions{sort&compress}

\journal{Journal of Magnetism and Magnetic Materials}

\DeclareMathOperator{\sign}{sign}

\begin{document}

\begin{frontmatter}

%opening
\title{Formation and growth of skyrmion crystal phase in a frustrated Heisenberg antiferromagnet with Dzyaloshinskii-Moriya interaction}
\author[1]{M. Mohylna},
\author[2,3]{J. Bu\v{s}a Jr.}
\author[1]{M. \v{Z}ukovi\v{c}\corref{cor1}}
\ead{milan.zukovic@upjs.sk}
\address[1]{Institute of Physics, Faculty of Science, P.J. \v{S}af\'arik University\\ Park Angelinum 9, 041 54 Ko\v{s}ice, Slovakia}
\address[2]{LIT JINR, Joliot-Curie 6, 141980 Dubna, Moscow region, Russia}
\address[3]{IEP SAS, Watsonova 47, 040 01 Košice, Slovakia}
%\maketitle
\cortext[cor1]{Corresponding author}

%\title{Formation and growth of skyrmion crystal phase in a frustrated Heisenberg antiferromagnet with Dzyaloshinskii-Moriya interaction}
%\author{M. Mohylna},
%\author{M. \v{Z}ukovi\v{c}}
%\email{milan.zukovic@upjs.sk}
%\affiliation{Institute of Physics, Faculty of Science, P.J. \v{S}af\'arik University, Park Angelinum 9, 041 54 Ko\v{s}ice, Slovakia}
%\date{\today}

\begin{abstract}
We study the formation and growth of a skyrmion crystal (SkX) phase in a frustrated antiferromagnetic Heisenberg model on a triangular lattice with Dzyaloshinskii-Moriya interaction (DMI) in the presence of an external magnetic field. We build phase diagrams in the temperature-field parameter plane, featuring first-order phase transitions and tricritical points, and study their evolution with the increasing DMI strength. It is found that already relatively small DMI intensity can lead to the appearance of the SkX phase at low temperatures in the vicinity of the meeting point of the remaining helical, coplanar up-up-down, and canted V-like ordered phases. By means of a parallel tempering (PT) Monte Carlo algorithm we find that the minimum value of DMI, at which the skyrmion phase emerges, $D_t \approx 0.02$, is one order of magnitude smaller than previously reported. We demonstrate the efficiency of PT in reliable and precise location of the phase boundaries between different phases separated by either very large (strong first-order transition) or barely detectable energy barriers, where the standard Monte Carlo approach may easily fail to find a stable solution.
\end{abstract}


%\PACS 05.50.+q \sep 64.60.De \sep 75.10.Hk \sep 75.30.Kz \sep 75.50.Ee \sep 75.50.Lk

\end{frontmatter}

\section{Introduction}

Nontrivial twisted magnetic configurations, called skyrmions, have recently drawn much attention due to their exotic properties~\cite{nagaosa}, which hold great potential for technological applications~\cite{spcur1,racetrack, spcur3, lg1, osc}. The possibility of the existence of such structures in magnetic materials was theoretically predicted more than 20 years ago~\cite{bogd1, bogd2, rossler} but the first direct experimental proof of the presence of the hexagonal skyrmion crystal phase in the small temperature-field window just below the Curie temperature was obtained in 2009 by M{\"u}hlbauer \emph{et~al.}~\cite{muhlbauer} in the MnSi bulk sample.

By now, many ferromagnets have been suggested as potential skyrmion hosting materials, though the creation and manipulation of the skyrmions in antiferromagnetic (AFM) materials~\cite{galkina2018dynamic,kravchuk2019spin, gorobets20203d} might have certain advantages~\cite{zhang_anti, barker, bessarab} over their ferromagnetic counterparts. For instance, skyrmions in AFMs are not subjected to the Magnus force~\cite{zhang_magnus}, which causes skyrmions' motion transversal to the current direction and can lead to their annihilation at the sample's edges. AFM skyrmions are formed from $n$ interpenetrating spin configurations following the $n$-sublattices structure of the AFM lattice. They have shown ability to overcome hole defects ~\cite{silva2019antiferromagnetic} and can be manipulated by temperature and anisotropy gradients in addition to the electric current~\cite{shen2018dynamics, khoshlahni2019ultrafast}. Furthermore, the Dzyaloshinskii-Moriya interaction (DMI), one of the mechanisms responsible for the stabilization of the skyrmions in magnetic materials~\cite{bogd2}, is more commonly found in AFM magnets rather than in ferromagnetic systems. Among other such mechanisms are frustrated exchange interactions~\cite{okubo}, four-spin exchange interactions~\cite{mono} or long-ranged dipolar interactions~\cite{dipol1, dipol2}, albeit the last ones lead to structures with unfixed helicity, thus, strictly speaking, they cannot be called topological magnetic skyrmions. On the other hand, the helicity of the skyrmions appearing due to the interplay of the exchange interaction, the DMI and the external magnetic field is uniquely set by the direction of the DMI vector, which, in turn, is defined by the geometry of the crystal and can be easily adjusted thanks to the modern engineering methods. The DMI can give rise to both N{\'e}el and Bloch type skyrmions of the size of 5-100 nm.

Recently it has been shown by Okubo \emph{et~al.}~\cite{okubo} that both skyrmion and antiskyrmion lattices can be formed on a triangular lattice with next-nearest interactions or any lattices with the trigonal symmetry due to the presence of frustration even in the absence of the DMI. Various new effects have been observed in the model with the frustration induced by inclusion of further-neighbor interactions in the presence of the single-ion anisotropy~\cite{leonov_15} and the competing DMI~\cite{leonov_19}. The experimental evidence of the presence of the skyrmion phase in centrosymmetric frustrated magnets in the absence of the DMI was obtained by Kurumaji \emph{et~al.}~\cite{x-ray}. Yu \emph{et~al.}~\cite{Shartr} demonstrated, that the combination of the DMI and the frustration widens the range of fields with stabilized skyrmion lattice and increases the skyrmion density. Rosales \emph{et~al.}~\cite{rosales} studied the appearance of three ferromagnetic sublattices in a frustrated antiferromagnetic classical Heisenberg model on the triangular lattice for a fixed value of the DMI and demonstrated, that the skyrmion phase is stable in a quite wide temperature and field range, contrary to the case of an unfrustrated antiferromagnetic square lattice. The corresponding quantum model, studied by using quantum Monte Carlo method, suggested the existence of the skyrmion phase even at relatively high temperatures~\cite{fr}. The connection between the skyrmions in the antiferromanetic triangular Heisenberg lattice with the DMI and $\mathbb{Z}_2$ vortices in the pure Heisenberg triangular antiferromagnet was discussed by Osorio \emph{et~al.}~\cite{osor}. The enhanced stability of the aniferromagnetic skyrmions on a square lattice with prolonged lifetime of the order of milliseconds, compared to the ferromagnetic ones, was shown by Bessarab \emph{et~al.}~\cite{bessarab}. A so-called pseudo-skyrmion phase was confirmed to exist in a highly frustrated antiferromagnetic kagome lattice~\cite{kagome}. Overall, the combination of the antisymmetric DMI interaction and geometrical frustration has proved to lead to the stabilization of highly non-trivial phases. 

For practical purposes an important task is to determine the conditions, under which the formation of skyrmion patterns becomes favorable and for which materials those conditions are the least demanding from the implementation point of view. In the present study we follow up on earlier work by Rosales \emph{et~al.}~\cite{rosales} with the aim to study the formation and growth of the skyrmion lattice in a frustrated antiferromagnetic triangular monolayer with Heisenberg exchange interaction and the DMI in the presence of an external magnetic field. In particular, we build phase diagrams in the temperature-field parameter plane and study their evolution with the increasing DMI strength. By means of the parallel tempering approach we find that the limiting bottom value of the DMI, for which the skyrmion phase is still stable, is much lower than previously established by the low-energy effective theory~\cite{osorio_17} and the standard Monte Carlo simulation~\cite{mohylna_20}. Furthermore, the parallel tempering proves to be more efficient and precise in locating the phase boundaries, particularly between the helical and skyrmion phases, in the temperature-field plane. 

\section{Model and Method}
\begin{figure}[t!]
 \centering % <-- added

  \includegraphics[scale=0.2,clip]{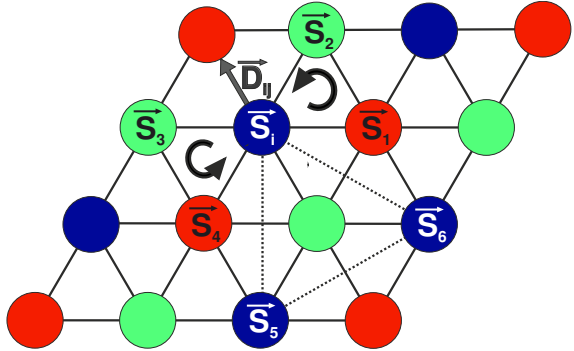} 
  \caption{Triangular lattice with the sites belonging to one of the three sublattices marked by different colors. $\vec{D}_{ij}$ is the DMI vector and arrows show the (counter-clockwise) direction of the calculation of local chirality and skyrmion number.}
  \label{latt}
\end{figure}
We consider the classical Heisenberg model on a triangular lattice with the Hamiltonian
\begin{equation}
\mathcal{H} = - J \sum_{\langle i,j \rangle}\vec{S_{i}}\cdot
\vec{S_{j}} + \sum_{\langle i,j \rangle}\vec{D_{ij}}
\cdot\Big [\vec{S_{i}}\times\vec{S_{j}} \Big] - h\sum_i S_i^z,
\label{hamilt}
\end{equation}
where $\vec{S_i}$ is a classical Heisenberg spin (vector of unit length) at the $i$th site, $J < 0$ is the antiferromagnetic exchange coupling constant (hereafter we put $J=-1$), $h$ is the external magnetic field applied perpendicular to the lattice plane (along the $z$ direction) and $\langle i,j\rangle$ denotes the sum over nearest-neighbor spins. $\vec{D}_{ij}$ is the Dzyaloshinskii-Moriya vector. In line with Ref.~\cite{rosales}, we chose it to point along the radius-vector $\vec{r}_{ij}=\vec{r}_i - \vec{r}_j$ connecting two neighboring sites, i.e., $\vec{D}_{ij} = D\frac{\vec{r}_{i,j}}{|\vec{r}_{i,j}|}$ (Fig.~\ref{latt}). We note that the DMI vector chosen along the bond direction rather than orthogonal to it, as the Moriya’s rule would dictate, gives rise to spiral spin states with the spins lying in a plane perpendicular to the propagation vector (as is experimentally the case for the chiral magnet MnSi), thus leading to the formation of the Bloch type skyrmions. The strength of the DMI is defined by the magnitude of the parameter $D$.

Our approach combines two Monte Carlo (MC) methods. The main technique is the parallel tempering (PT) or replica exchange MC~\cite{pt}, which has proved to be a powerful tool in the study of the systems with a complex energy surface, prone to get stuck in local minima. We run simulations for the linear lattice sizes $L = 24 - 120$, using $180-220$ replicas (temperatures), depending on the system size and the magnitude of the DMI. The temperature set is constructed by manually fine-tuning the geometrical progression to ensure adequate exchange rates at low temperatures and reasonable resolution at higher ones. The simulations are massively parallelized by implementing them on General Purpose Graphical Processing Units (GPGPU) using CUDA, which allowed to simulate all the replicas at different field values simultaneously. For each replica we use $2-4 \times 10^6$ MC sweeps for equilibration and the same amount for calculating mean values. The replica swapping occurs after each Metropolis sweep through the whole lattice. To identify first-order phase transitions, accompanied with hysteretic behavior of various quantities, and to approximately locate tricritical points, we also apply a hybrid MC (HMC) method, which combines the standard Metropolis algorithm with the over-relaxation method~\cite{or}. The latter is a deterministic energy preserving perturbation method, which helps decorrelate the system and leads to faster relaxation. In HMC we use up to $8 \times 10^6$ MC sweeps. Periodic boundary conditions were implemented in all the simulations.

To identify different phases and determine the phase boundaries we calculate several basic quantities, such as the magnetization $m$, the magnetic susceptibility $\chi_m$, and the heat capacity $C_v$, as follows:

\begin{equation}
m = \frac{\langle M \rangle}{N} = \frac{1}{N} \Big\langle \sum_i S_i^z\Big\rangle,
\label{magn}
\end{equation}

\begin{equation}
\chi_m = \frac{\langle M^2\rangle  - \langle M \rangle^2 }{NT},
\label{magnsus}
\end{equation}

\begin{equation}
C_v = \frac{\langle \mathcal{H}^2\rangle  - \langle \mathcal{H} \rangle^2 }{NT^2}, 
\label{heatcap}
\end{equation}
where $N = L^2$ is the total number of the lattice sites.

In addition to the standard thermodynamic quantities we also compute the skyrmion chirality and the skyrmion number, the discretizations of a continuum topological charge~\cite{berg}, which reflect the number and the nature of topological objects present in the system. The topological charge of a single skyrmion is $\pm 1$ for the core magnetization $\pm S$~\cite{obzor}. The skyrmion chirality $\kappa$, the skyrmion number $q$, and the corresponding susceptibilities $\chi_{\kappa}$ and $\chi_{q}$ are defined as follows:

\begin{equation}
\kappa = \frac{\langle K \rangle}{N} = \frac{1}{8\pi N} \Big\langle \sum_i \Big( \kappa^{12}_{i} + \kappa^{34}_{i} \Big)\Big\rangle,
\label{chiral}
\end{equation}

\begin{equation}
q = \frac{\langle Q \rangle}{N_s} = \frac{1}{4\pi N_s} \Big\langle\Big|\sum_i \Big( A^{12}_{i} \sign(\kappa^{12}_{i}) +A^{34}_{i} \sign( \kappa^{34}_{i}) \Big)\Big|\Big\rangle,
\label{sknum}
\end{equation}

\begin{equation}
\chi_{\kappa}= \frac{\langle K^2\rangle  - \langle K \rangle^2 }{NT},  
\label{xisus}
\end{equation}

\begin{equation}
\chi_q = \frac{\langle Q^2\rangle  - \langle Q \rangle^2 }{N_sT},
\label{xiQsus}
\end{equation}
where $\kappa^{ab}_{i} = \vec{S_i}\cdot[\vec{S_a}\times\vec{S_b}]$ is a chirality of a triangular plaquette of three neighboring spins ($\{\vec{S_i}, \vec{S_1}, \vec{S_2}\}$ and $\{\vec{S_i}, \vec{S_3}, \vec{S_4}\}$ in Fig.~\ref{latt} taken in the counter-clockwise fashion) and $A^{ab}_{i} = ||(\vec{S_a} - \vec{S_i})\times(\vec{S_b} - \vec{S_i})||/2$ is the area of the triangle spanned by those spins. The chirality is calculated for the whole lattice and the summation runs through all the spins, whereas the skyrmion number is calculated for each of the three sublattices, hence $N_s$ in Eqs. (\ref{sknum}) and (\ref{xiQsus}) is the number of sites in each of the sublattices, $N_s = L^2/3$, and the triangular plaquette for the local quantities is formed by the neighboring spins of the given sublattice ($\{\vec{S_i}, \vec{S_5}, \vec{S_6}\}$ in Fig.~\ref{latt}).

\section{Results}
\subsection{Skyrmion lattice formation}
The phase diagram in the absence of the DMI has been studied by several authors~\cite{kawamura_85,gvozdikova_2011,seabra_2011}. In zero field the ground state of the model is known to be ordered in a $120^{\circ}$ three-sublattice structure described by the
wave vector $\vec{k} = (4\pi/3,0)$. In a relatively small magnetic field $h < 3$ the $120^{\circ}$ becomes canted in a coplanar Y state, with one spin pinned in the negative $z$ direction and two canting up. At exactly $h = 3$ the spins order in a collinear state, with two spins up and one spin down (UUD), resulting in a
$1/3$ magnetization plateau. For $h>3$ up to the fully polarized paramagnetic state, reached at the saturation field $h=9$, a coplanar V canted state fulfilling the constraint $\vec{S_1} + \vec{S_2} + \vec{S_3} = \vec{h}/3$ on each plaquette becomes preferred. The fact that the orientation of the spin plane and sublattice directions inside the plane remain undetermined results in degeneracy, which is however lifted by thermal fluctuations due the order-from-disorder effect~\cite{kawamura_1984}. The resulting phase diagram in $T-h$ plane can be found in Refs.~\cite{kawamura_85,gvozdikova_2011,seabra_2011}.

Recent investigations by the low-energy effective theory~\cite{osorio_17} and the standard Monte Carlo simulation~\cite{mohylna_20} suggested that the skyrmion lattice (SkX) phase emerges at very low temperatures in a narrow field range around $h=3$, providing that the DMI intensity exceeds the threshold value $D_t \approx 0.2$. Bellow we show that by employing the PT method we are able to detect the SkX phase at much smaller DMI and by performing some finite size analysis we reestimate $D_t$ to be even one order smaller than previously reported. 

\begin{figure}[t!]
\centering
\vspace{-5mm}
\subfigure{\includegraphics[scale=0.32,clip]{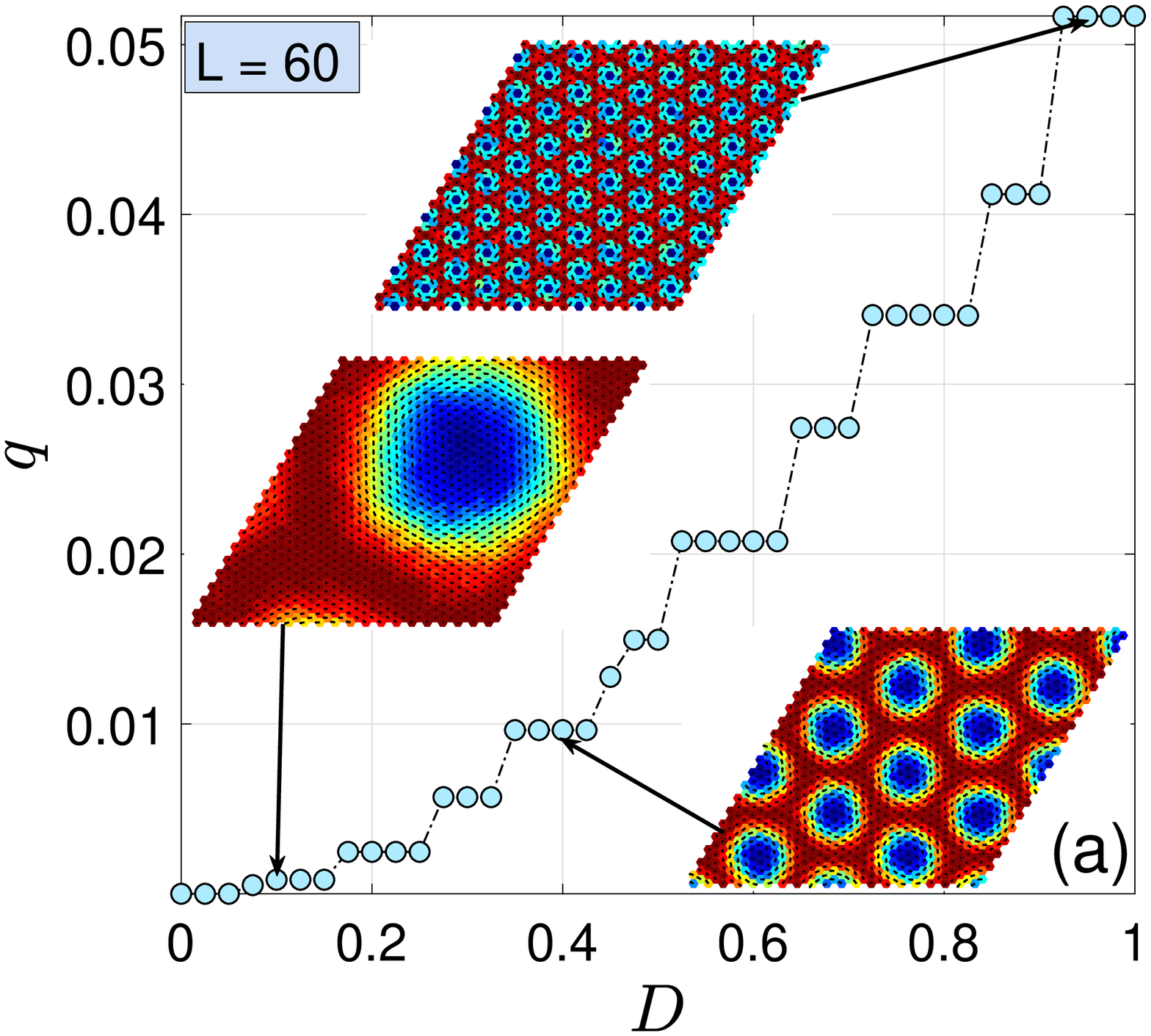}\label{fig:H3_2}}
\subfigure{\includegraphics[scale=0.32,clip]{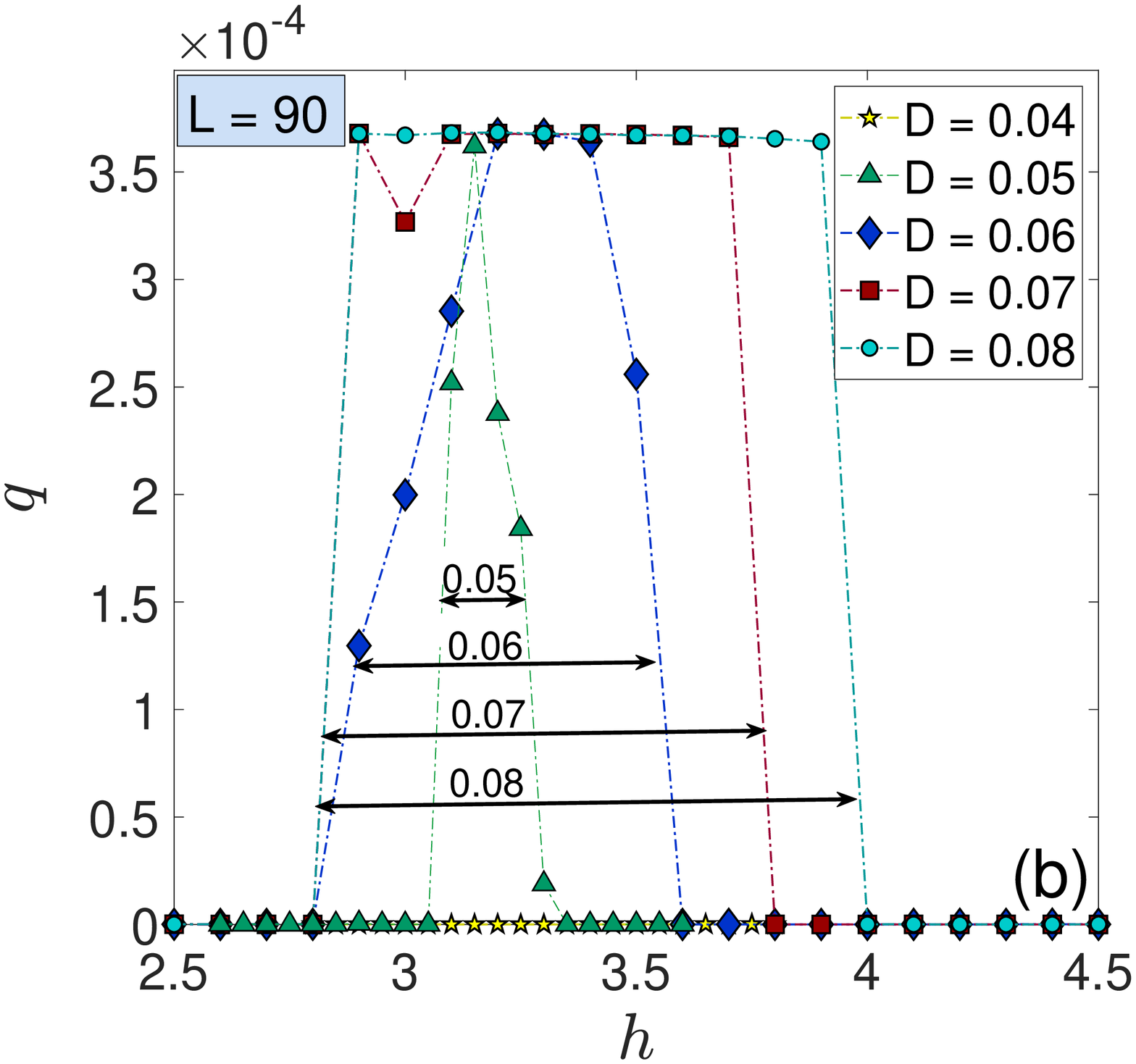}\label{fig:xiQL90}}\\ \vspace{-5mm}
%\subfigure{\includegraphics[scale=0.25,clip]{L90D010.eps}\label{fig:xi-m}}
%\subfigure{\includegraphics[scale=0.25,clip]{L120D010.eps}\label{fig:xi-m}}
%\subfigure{\includegraphics[scale=0.25,clip]{L120D006.eps}\label{fig:xi-m}}\\
\subfigure{\includegraphics[scale=0.32,clip]{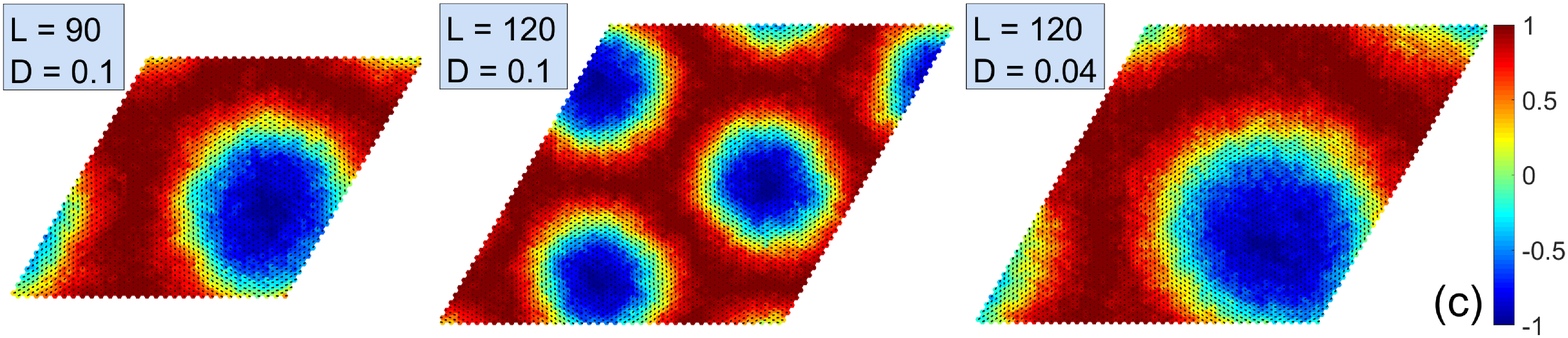}\label{fig:snapsh}} \\ \vspace{-5mm}
\subfigure{\includegraphics[scale=0.32,clip]{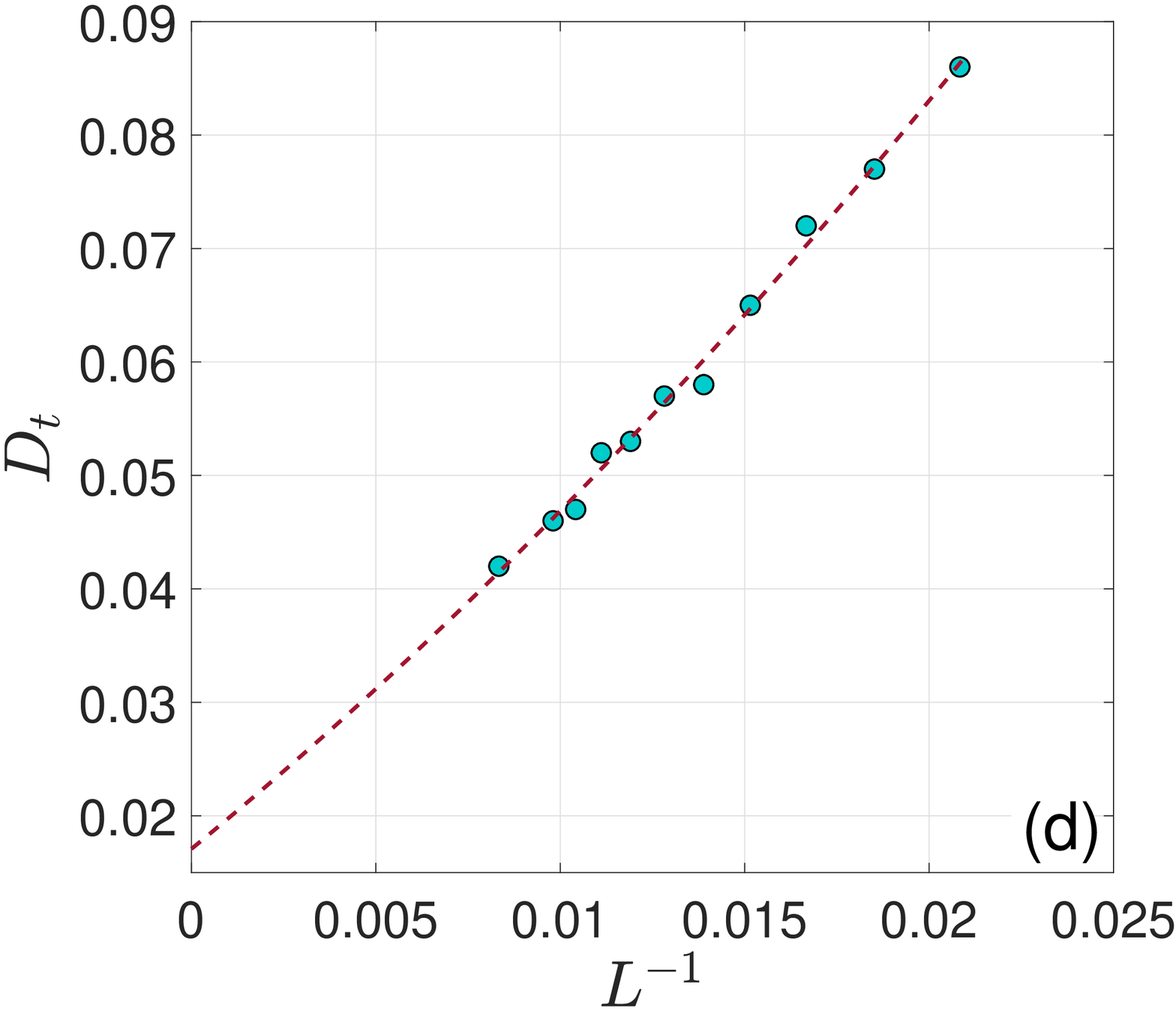}\label{fig:latDep}} \vspace{-3mm}
\caption{(a) Dependence of the skyrmion number on $D$ for $h = 3$ and $L = 60$. (b) Field dependencies of the skyrmion number for $L = 90$ and several small values of $D$, close to the threshold $D_t(L=90) \approx 0.052$. (c) Snapshots of skyrmions on one sublattice for $h = 3.2$ and (left) $L = 90$, $D = 0.1$, (middle) $L = 120$, $D = 0.1$ and (right) $L = 120$, $D = 0.042$. (d) Threshold values of $D_t(L)$ extrapolated to $D_t(L \to \infty) \approx 0.02$. All results are presented for $T = 0.009$.}
\label{lowD}
\end{figure}

In Fig.~\ref{fig:H3_2} we show the dependence of the skyrmion number on the magnitude of the DMI for a moderate lattice size $L = 60$ in the region where, based on the previous works, the SkX phase is expected to emerge, i.e., low temperature ($T = 0.009$) and $h = 3$. The result is a step-wise increasing curve with the first non-zero plateau emerging at $D \approx 0.075$. The spin snapshot on one of the three sublattices in the inset, taken at $D = 0.1$, demonstrates that already the first non-zero plateau corresponds to the SkX phase, with one huge skyrmion filling the entire sublattice. We note that this complex antiferromagnetic SkX phase is made up from three interpenetrating skyrmion lattices, one for each sublattice, and thus similar snapshots can be observed also on the remaining two sublattices. As the DMI increases, the size of the skyrmions becomes smaller and thus more and more of them can fit into the sublattice (see snapshot in the inset at $D = 0.4$). However, the skyrmion lattice, which in the present system corresponds to a close packing of individual skyrmions on a triangular lattice, has to be commensurate with the sample size, which is fixed. This translates to the increase of the skyrmion number in a step-wise fashion, until the structure consisting of one central spin pointing opposite to the field direction and six surrounding in-plane spins is formed (see snapshot in the inset at $D = 0.95$). 

We found the threshold DMI strength for $L = 60$ to be around $D_t = 0.072$. However, it turns out to be lattice size dependent and as we increase the lattice size to $L = 90$ (Fig.~\ref{fig:xiQL90}), $D_t$ shifts to even lower values. For $D = 0.08$ the SkX phase is present in a surprisingly wide field range $2.8 \leq h < 4$ but as the DMI approaches its limiting value $D_t(L=90) \approx 0.052$ the SkX phase appears only in a close vicinity of $h = 3.2$. For $L = 90$ the SkX phase at low $D$ is represented by a single giant skyrmion filling the whole sublattice (see the first snapshot in Fig.~\ref{fig:snapsh}). However, as we further increase the lattice size to $L = 120$ with the fixed $D = 0.1$, multiple skyrmions on a regular triangular lattice pattern appear (the second snapshot in Fig.~\ref{fig:snapsh}). Further decrease of $D$ reduces it to one giant skyrmion again (the third snapshot in Fig.~\ref{fig:snapsh}) and such a state persists down to $D_t(L=120)=0.042$, which is lower then $D_t(L = 90)$. 

Based on the above observation, in Fig.~\ref{fig:latDep} we performed a simple finite-size analysis by plotting the values $D_t$ estimated for different lattice sizes and extrapolating them to infinity. Our analysis suggests that in the thermodynamic limit the SkX phase should appear already at $D_t(L \to \infty) \approx 0.02$, which is one order of magnitude smaller than the earlier estimates.

\subsection{Phase diagrams}
\subsubsection{Thermodynamic quantities}

\begin{figure}[t!]
\centering
\subfigure{\includegraphics[scale=0.32,clip]{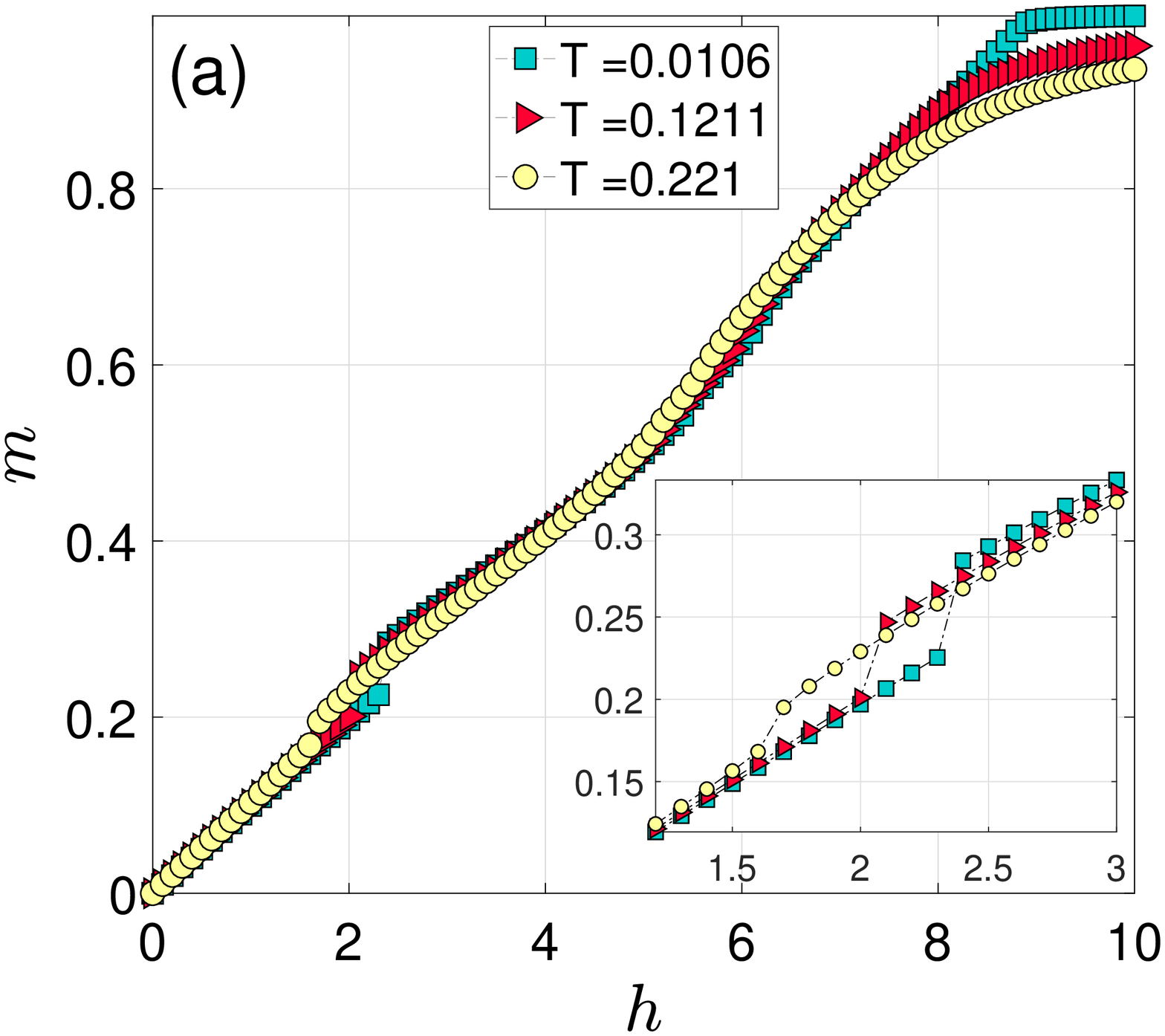}\label{fig:h-m}}
\subfigure{\includegraphics[scale=0.32,clip]{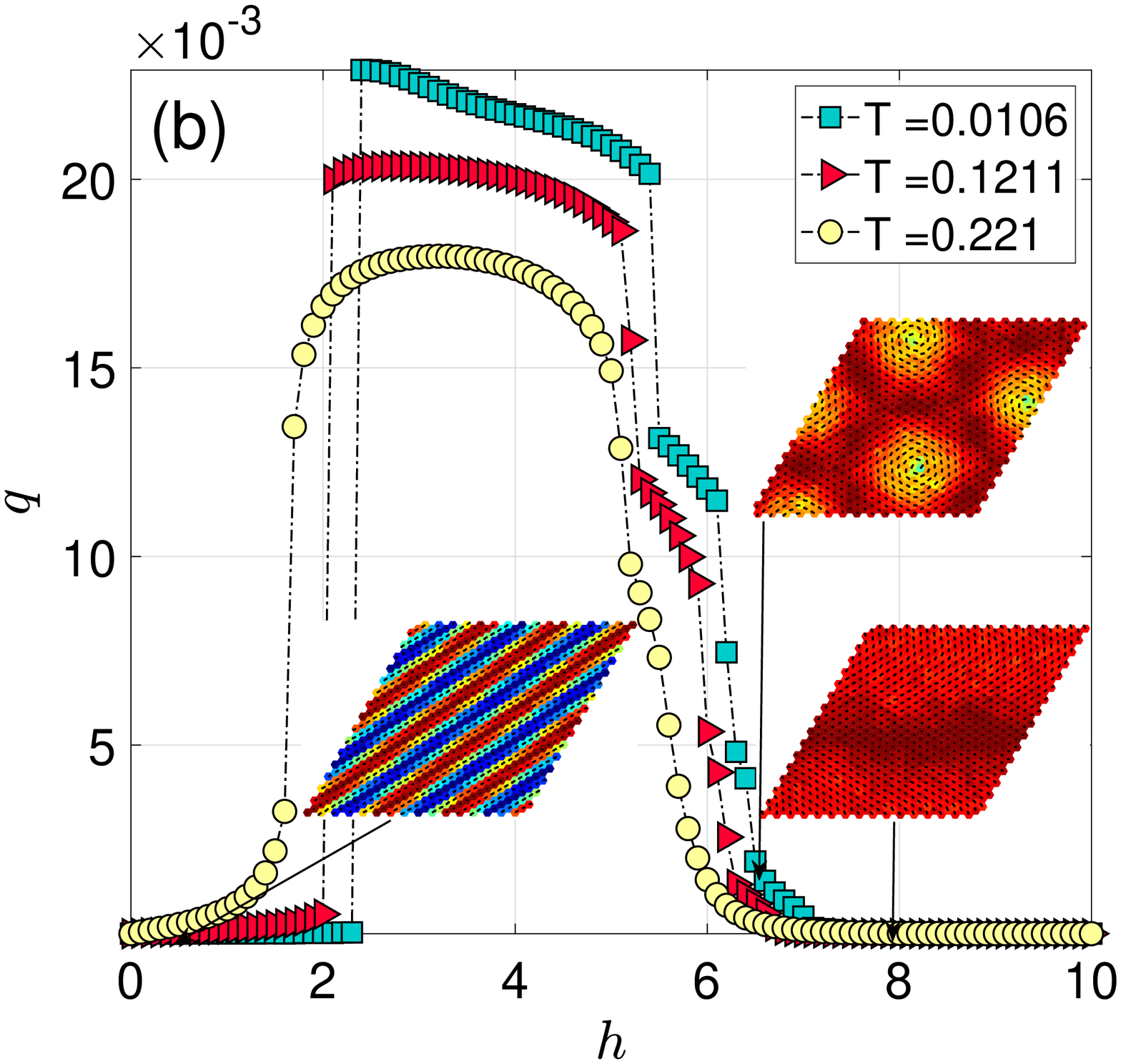}\label{fig:Q-m}}
\subfigure{\includegraphics[scale=0.32,clip]{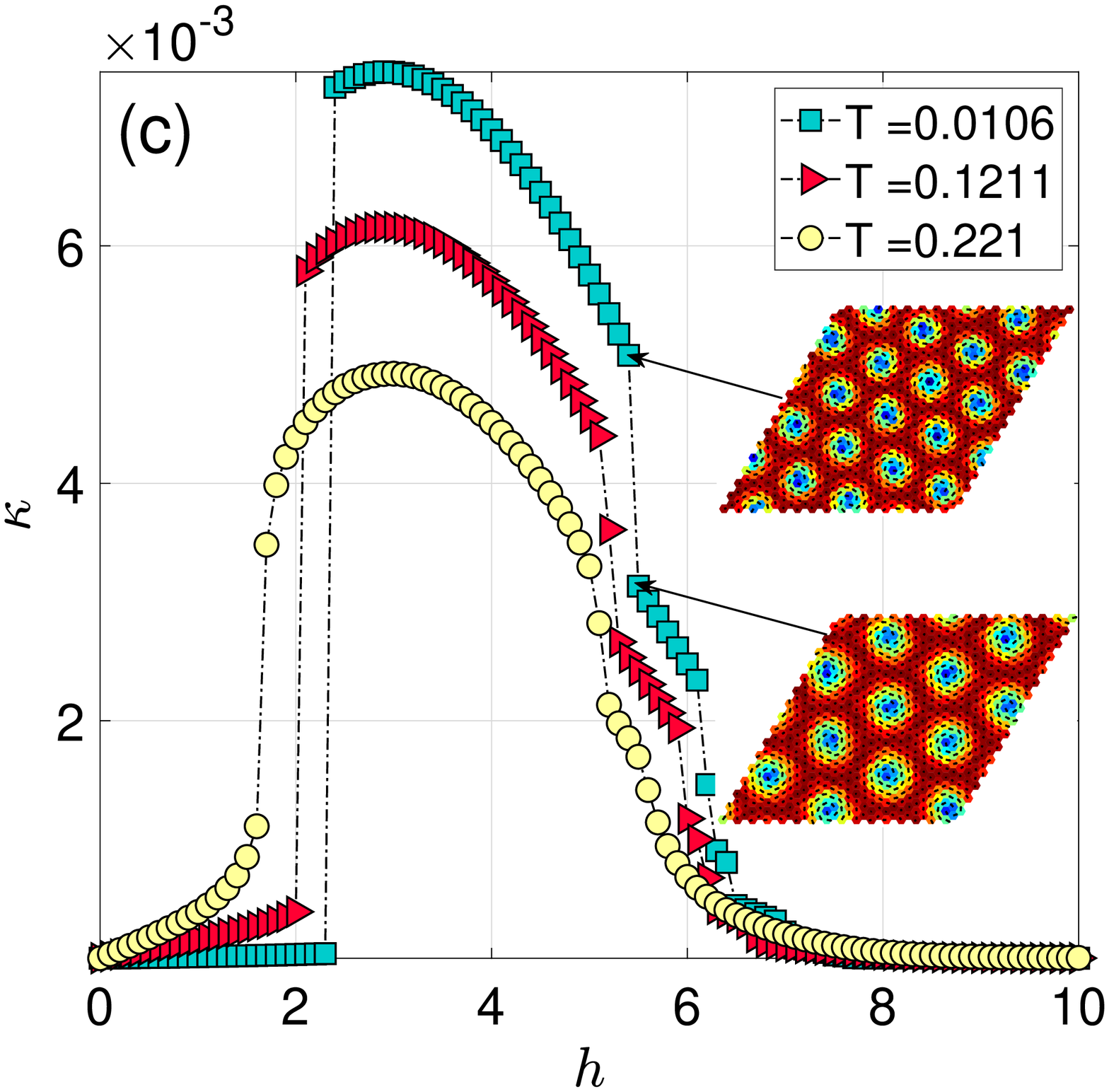}\label{fig:xi-m}}
\subfigure{\includegraphics[scale=0.32,clip]{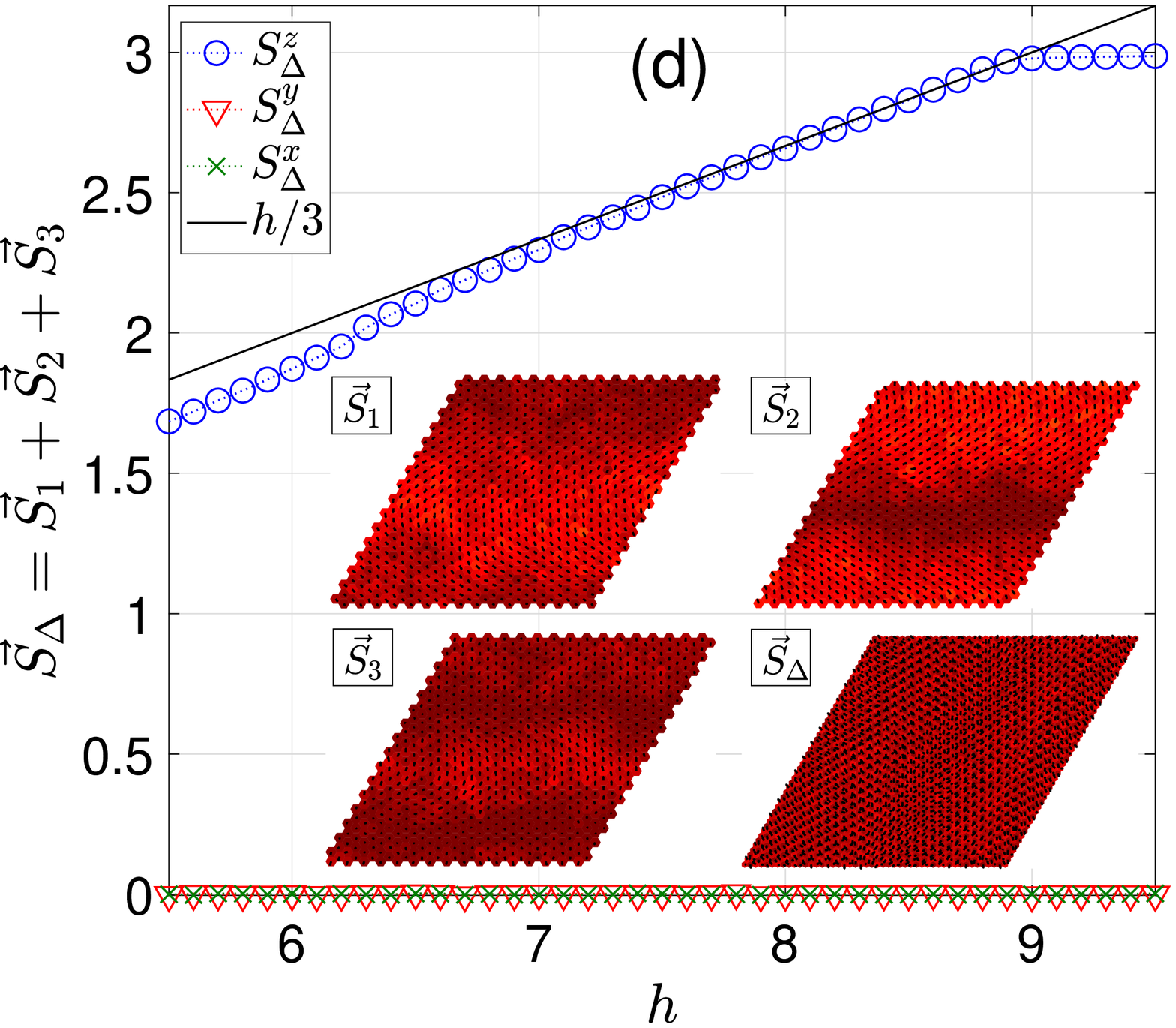}\label{fig:S-m}}
\caption{Field dependencies of (a) the magnetization, (b) the skyrmion number, and (c) the chirality, for selected temperatures, $L = 48$, and $D = 0.6$. The inset in (a) zooms on the magnetization process at the discontinuous HL-SkX phase transition, the inset in (b) shows one sublattice snapshots within the HL ($h=0.5$) and VL ($h=6.5$ and $h=7.8$) phases, and the inset in (c) shows sublattice snapshots at two neighboring field values separated by the chirality jump at the SkX-VL transition. Panel (d) shows field dependencies of the components of the vector $\vec{S}_{\Delta}=\vec{S_1} + \vec{S_2} + \vec{S_3}$ averaged over all triangular plaquettes at $T=0.0106$, as well as snapshots of spins $\vec{S_1}$, $\vec{S_2}$, and $\vec{S_3}$ on all three sublattices along with the composite spins $\vec{S}_{\Delta}$, taken at $h=7.8$.}
\label{h_dep}
\end{figure}

Below we demonstrate the behavior of the calculated thermodynamic quantities in the $T-h$ parameter space, for the fixed values of $D = 0.6$ and $L = 48$. In the low-temperature region different phases and phase boundaries between them can be conveniently determined based on the magnetization and the skyrmion chirality or the skyrmion number. The latter two are the order parameters of the SkX phase and thus usable for identification of the boundaries between the SkX and the neighboring phases, while the saturation value of the former signals the crossover to the fully polarized paramagnetic (P) phase. 

The field dependencies of the respective quantities for several temperature values are plotted in Fig.~\ref{h_dep}. In the magnetization process (Fig.~\ref{fig:h-m}) one can observe several more or less visible anomalies, related to the successive phase transitions from the helical phase (HL) to SkX, from SkX to the V-like (VL) (see its description below), and finally from VL to the P phase. Probably the most apparent is the magnetization jump at $h \approx 2$, related to the onset of the SkX phase, as also evidenced in Figs.~\ref{fig:Q-m} and \ref{fig:xi-m} showing the order parameters. The discontinuous character of the change of all the quantities points to the first-order phase transition, in accordance with the conclusions of the previous work~\cite{rosales}. The increasing temperature slightly smoothens out their sharply discontinuous behavior and shifts the transition boundary to smaller field values.

Less conspicuous but multiple anomalies in the magnetization curve can be observed at $h \approx 6$, where the system crosses from the SkX to the VL phase. The transition is accompanied by still sharp but step-wise decreasing SkX order parameters (Figs.~\ref{fig:Q-m} and \ref{fig:xi-m}). However, unlike in the HL-SkX transition, we believe that the multiple discontinuities in the SkX-VL phase transition do not signify its first-order character but appear just as an artifact of the finiteness of the lattice. The lattice size is fixed and can accommodate only certain number of skyrmions of a given radius. The increasing field tends to increase the magnetization by making the radius of skyrmions grow until they do not fit into the lattice and further growth can only be facilitated by reducing their number (see snapshots in the inset of Fig.~\ref{fig:xi-m}). Thus, with the increasing $L$ one should expect a larger number of smaller steps and eventually convergence to a continuous behavior in the thermodynamic limit. Phenomenological description of this phenomenon, as the competition between the applied magnetic field and the DMI, was provided by Rosales \emph{et~al.}~\cite{rosales}. 

Upon further field increase, the spin texture changes. The snapshot of spins on one sublattice, taken close to the SkX-VL phase boundary at $h=6.5$ (see the inset of Fig.~\ref{fig:Q-m}), shows that spins still form vortical-like structure, similar to the SkX phase, but with all $z$-components positive. However, further away from the phase boundary at higher fields the spin texture resembles wave-like pattern with a rather strong alignment of the $z$-components with the field direction (see the snapshot in the inset of Figs.~\ref{fig:Q-m} and~\ref{fig:S-m} taken at $h=7.8$). Thus, within the VL phase the sublattice uniformity of the coplanar V phase at $D=0$ is lost. Nevertheless, the constraint $\vec{S_1} + \vec{S_2} + \vec{S_3} = \vec{h}/3$ appears to be obeyed on each plaquette. This is demonstrated in Fig.~\ref{fig:S-m}, which shows field dependencies of the components of the vector $\vec{S}_{\Delta}=\vec{S_1} + \vec{S_2} + \vec{S_3}$ averaged over all triangular plaquettes as well as snapshots of spins $\vec{S_1}$, $\vec{S_2}$, and $\vec{S_3}$ on all three sublattices along with the composite spins $\vec{S}_{\Delta}$ for $T=0.0106$. One can see that within the VL phase the vectors $\vec{S}_{\Delta}$ are aligned with the field direction and their magnitude increases as $h/3$. The small deviation at lower fields might be caused by thermal fluctuations, since the results were obtained at finite temperature. The uniformity of the snapshot $\vec{S}_{\Delta}$ shows that the constraint is fulfilled not only on average but also locally on each plaquette. At around $h=9$, the lowest-temperature magnetization curve reaches its saturation point, signaling the crossover to the fully polarized P phase.

\begin{figure}[t!]
\centering
\subfigure{\includegraphics[scale=0.32,clip]{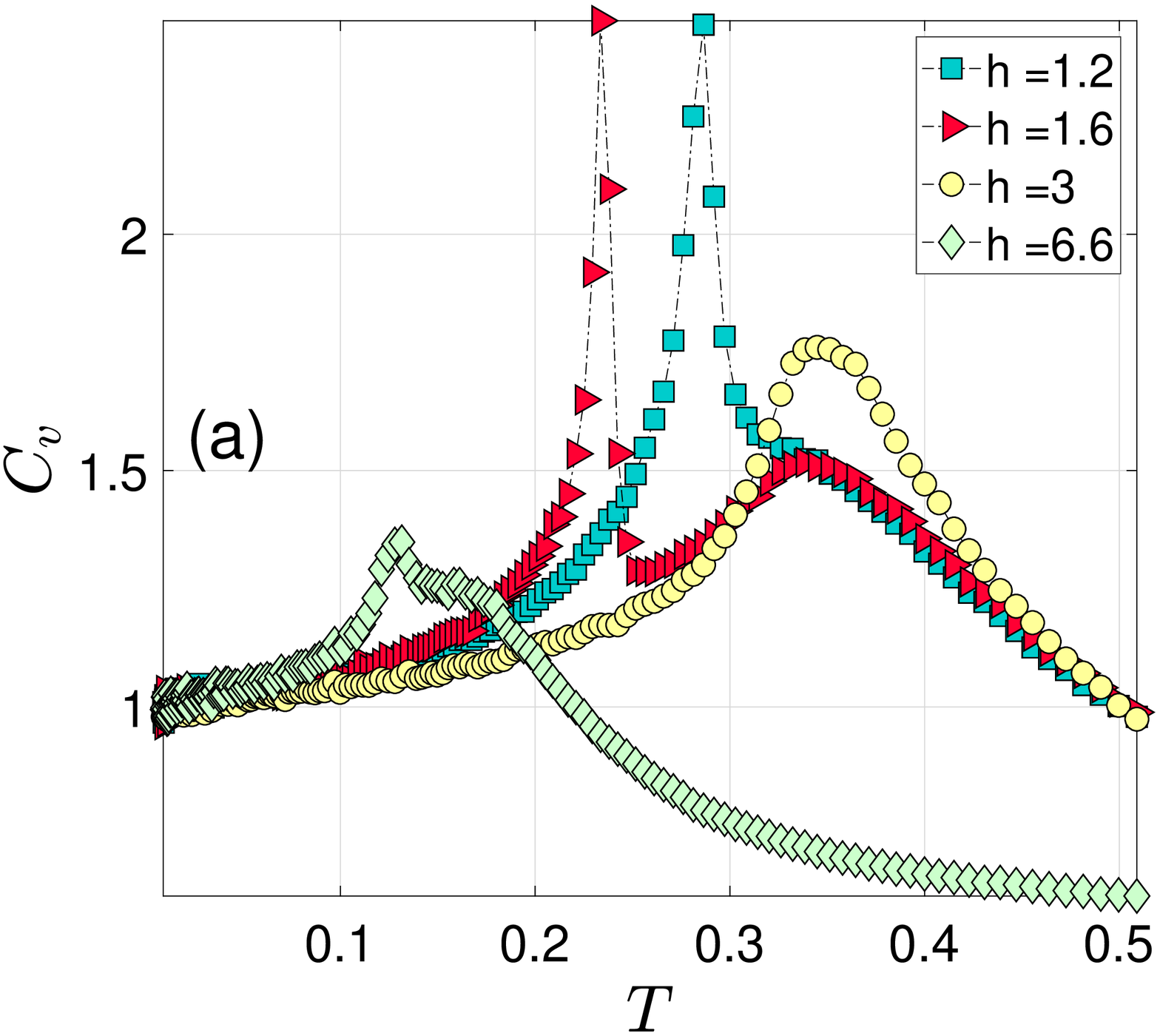}\label{fig:T-c}}
\subfigure{\includegraphics[scale=0.32,clip]{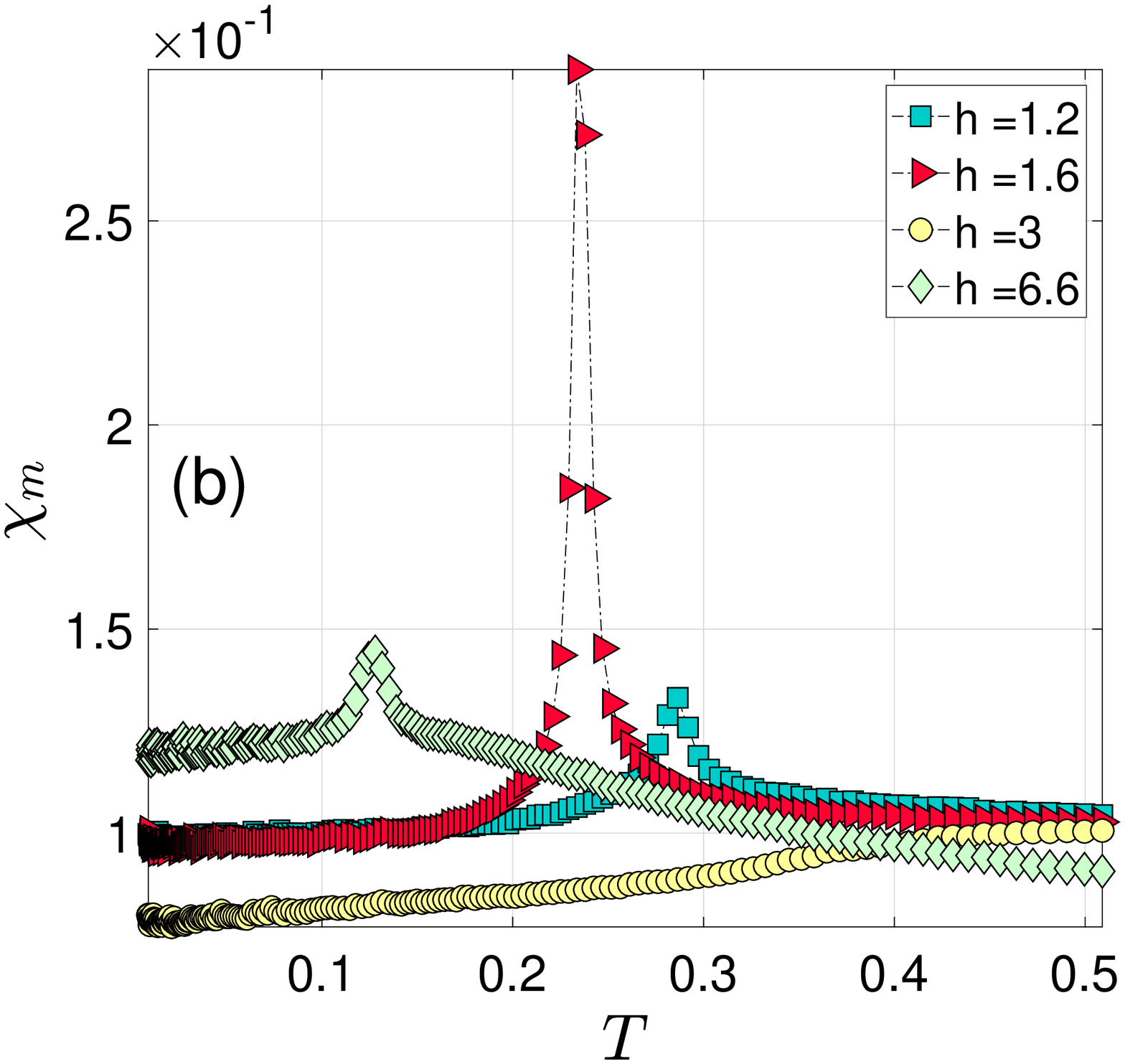}\label{fig:T-chi_m}}\\
\subfigure{\includegraphics[scale=0.32,clip]{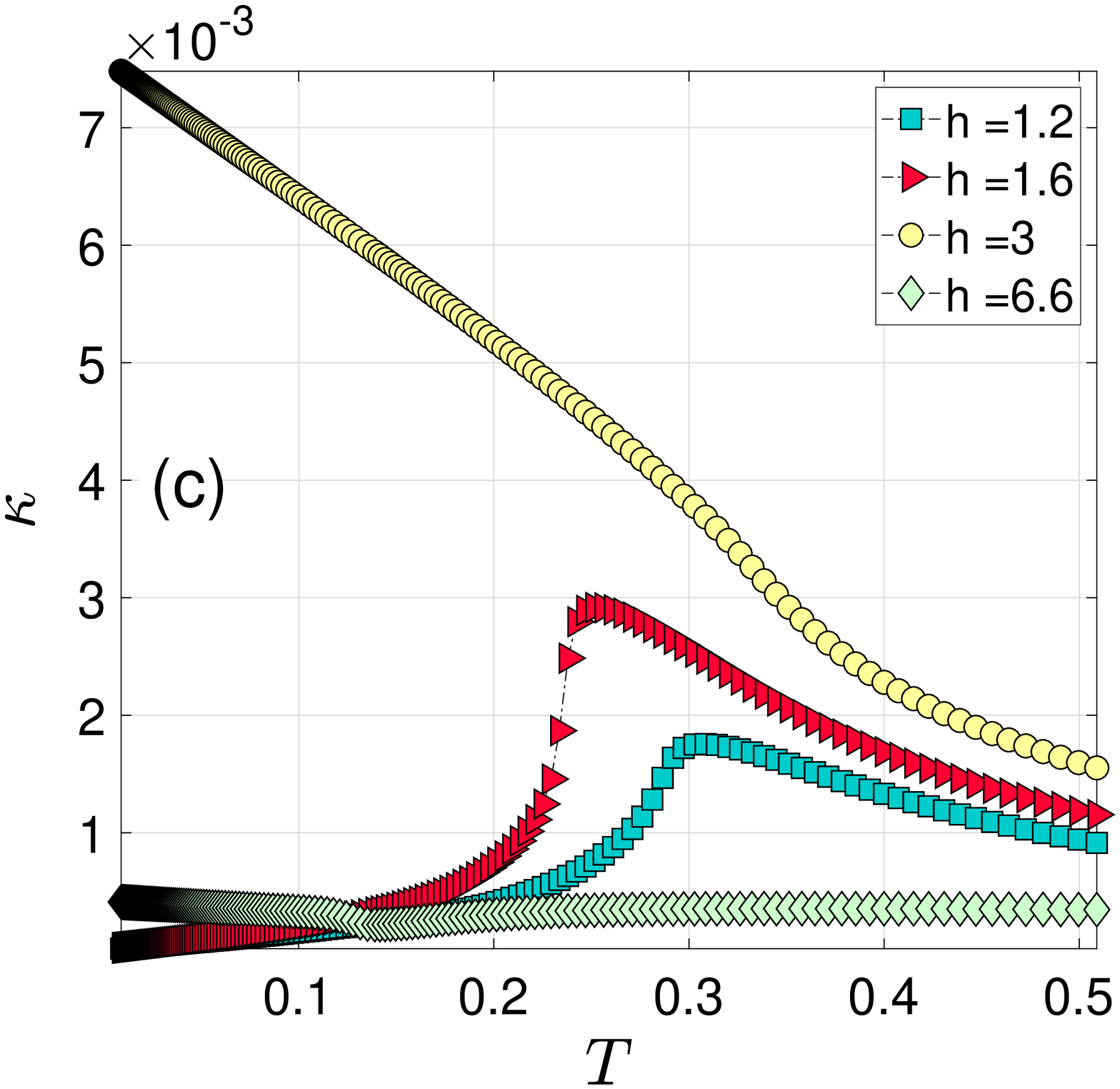}\label{fig:T-xi}}
\subfigure{\includegraphics[scale=0.32,clip]{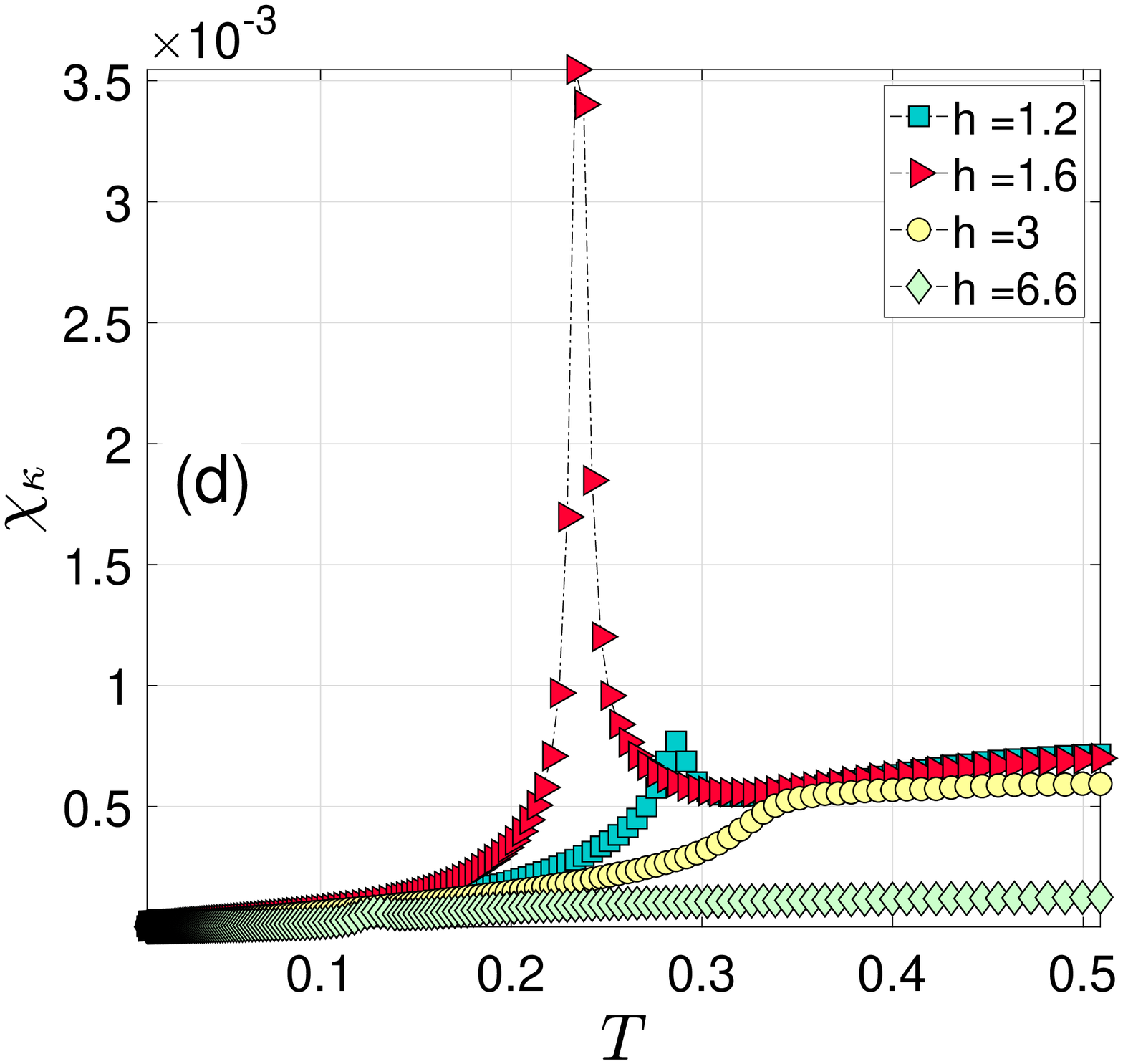}\label{fig:T-chi_xi}}
\caption{Temperature dependencies of (a) the specific heat, (b) the magnetic susceptibility, (c) the chirality, and (d) the chiral susceptibility, for selected fields, $L = 48$, and $D = 0.6$.}
\label{t_dep}
\end{figure}

For temperature dependencies we chose four representative field values corresponding to four types of the ordered phases: HL, SkX, VL and P and instead of the quantities presented in Fig.~\ref{h_dep}, we focus more on the response functions. The latter are more convenient for the phase boundary location at higher temperatures, where their peaks are broader and better detectable. On the other hand, at very low temperatures and particularly at the first-order phase transitions they may become too narrow (spike-like) to be captured at the given temperature resolution. 

For the smallest considered field $h = 1.2$ the single peak in all the response functions, shown in Fig.~\ref{t_dep}, suggests that the system undergoes only one phase transition at around $T = 0.29$ from the HL to the P state. The chirality (and the skyrmion number) is zero within the HL phase and has a small residual value in the disordered state due to the finiteness of the system. For $h = 1.6$ the skyrmion phase becomes stable in a small temperature window between the HL and P states. The sharp peaks in the heat capacity and both magnetic and chiral susceptibilities at $T \approx 2.3$ correspond to the HL-SkX transition and the smaller secondary peak in the heat capacity at $T \approx 0.34$ corresponds to another transition to the P phase. Further increase of the field leads to the disappearance of the HL phase and the SkX phase becomes stable in a quite wide temperature range (up to $T \approx 0.35$ for $h = 3$). Finally, for $h = 6.6$ only the VL phase is present and the transition to the P phase occurs at $T \approx 0.13$.

%%%%%%%%%%%% Hyster
\begin{figure}[t!]
 \centering % <-- added
  \includegraphics[scale=0.35,clip]{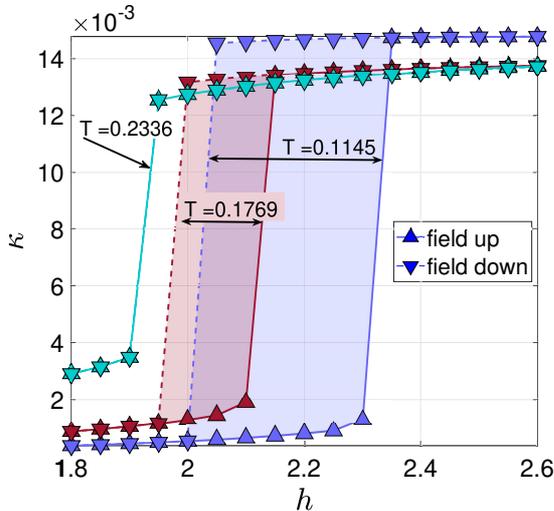} 
  \caption{Hysteresis loops in the chirality, obtained at different temperatures, $L = 48$ and $D = 1.0$, by gradually increasing the field from $h = 0$ (triangle-up solid lines) and decreasing it from $h = 10$ (triangle-down dashed lines). The double arrow lines show the shaded hysteresis widths (zero width at $T=0.2336$).}
	\label{hyster}
\end{figure}

As described above, the HL-SkX phase transitions at sufficiently low temperatures show signs of a discontinuous first-order character. In such a case standard MC simulations typically show hysteretic behavior of some quantities, such as the magnetization, when the transition is approached from low (high) fields by gradual increase (decrease) of the field intensity. It disappears when temperature approaches a tricritical point at which the character of the transition changes to the continuous second-order one. Therefore, monitoring the presence/absence of hystereses in the measured quantities can serve as an approximate method for the tricritical point location. In our HMC simulations such a hysteretic behavior, besides the magnetization, was also observed in the energy, the skyrmion number as well as the chirality. The last one is shown in Fig.~\ref{hyster}, for $L = 48$, $D = 1.0$ and three selected temperatures. For each temperature there are two curves: one is obtained by starting from a random initialization at $h = 0$ and increasing the external magnetic field up to $h = 10$ (triangle-up solid lines) and the other one by starting from the fully polarized state at $h = 10$ and decreasing it down to $h = 0$ (triangle-down dashed lines). As the temperature is increased the hysteresis width decreases until the two curves merge at $T=0.2336$ (green line). We note that the hysteresis width also shrinks with the increasing simulation time\footnote{The true boundary lies inside the hysteresis loop and could be obtained by taking the simulation time to infinity or applying a more sophisticated method, such as the parallel tempering.} and thus it can go to zero even though the transition is still first-order. This apparently happens also in our case, with relatively long simulation runs ($8 \times 10^6$ MC sweeps), when at $T=0.2336$ the hysteresis has disappeared but the discontinuity is still rather obvious. Consequently, one should be aware that this approach may lead to some underestimation of the true tricritical temperatures.

\subsubsection{Phase diagram evolution}

%%%%%%%%%%%%%%%%%%%phase diagrams

\begin{figure}[t!]
\centering
\subfigure{\includegraphics[scale=0.32,clip]{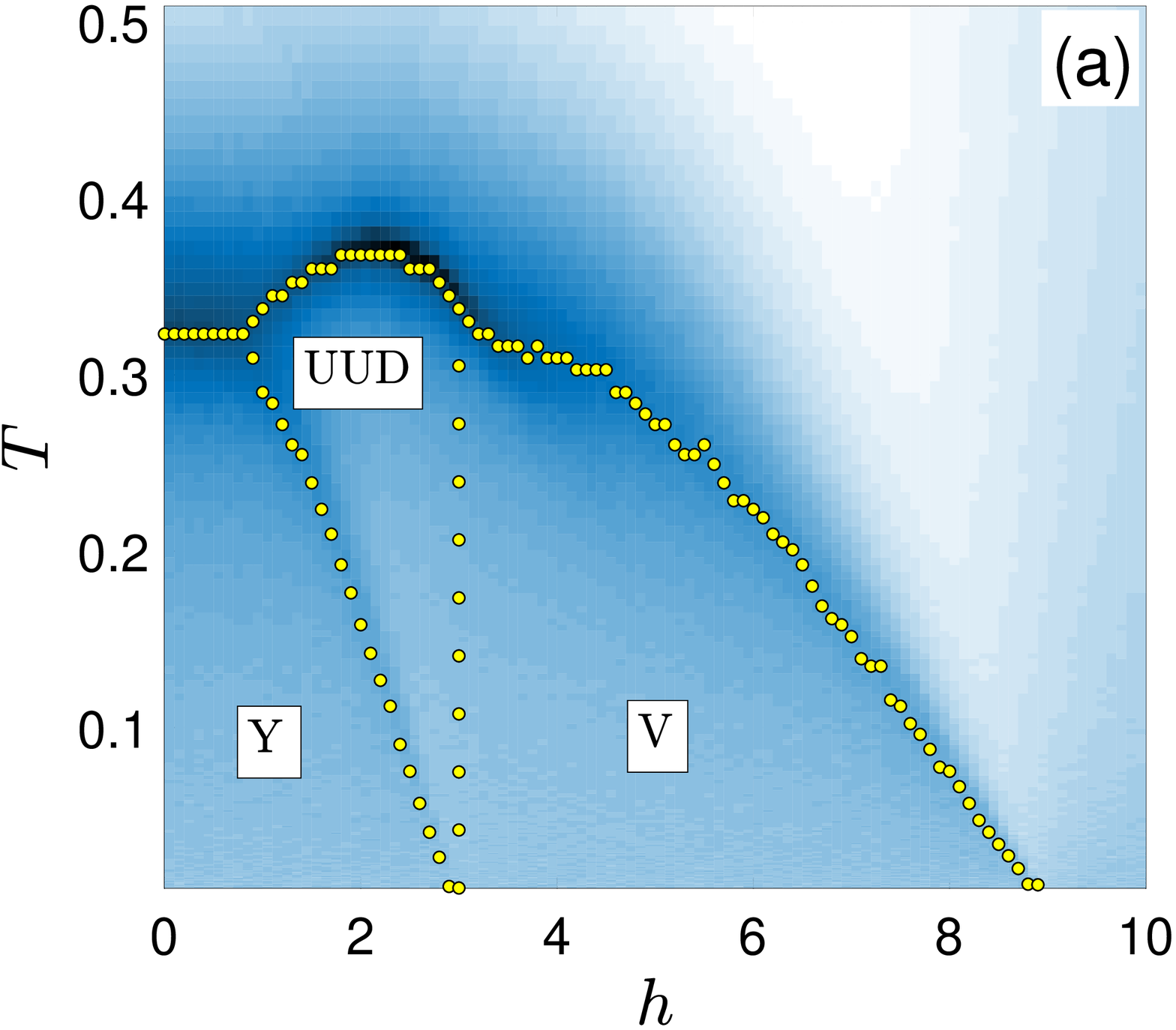}\label{fig:D000}}
\subfigure{\includegraphics[scale=0.32,clip]{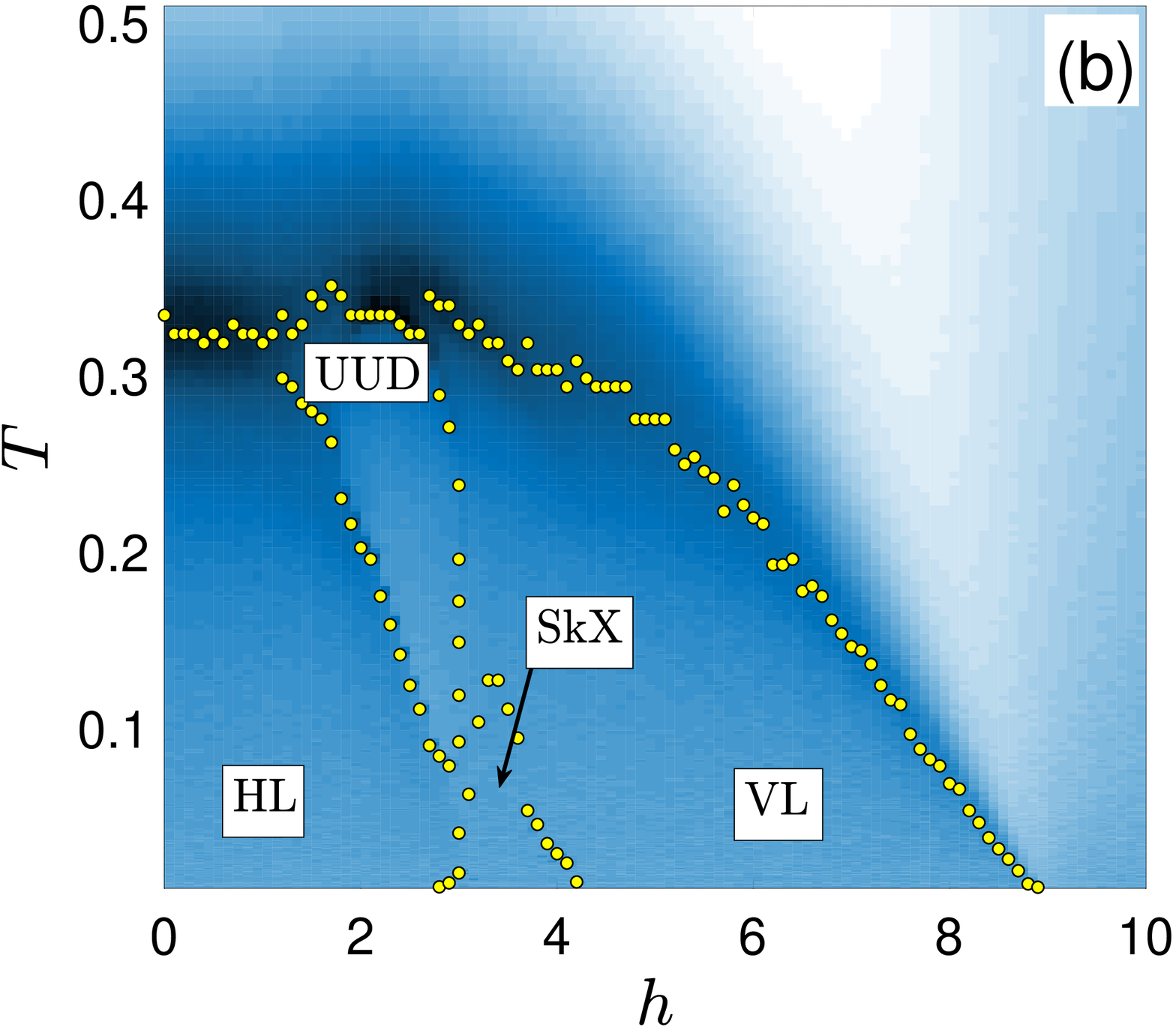}\label{fig:D010}}\\
\subfigure{\includegraphics[scale=0.32,clip]{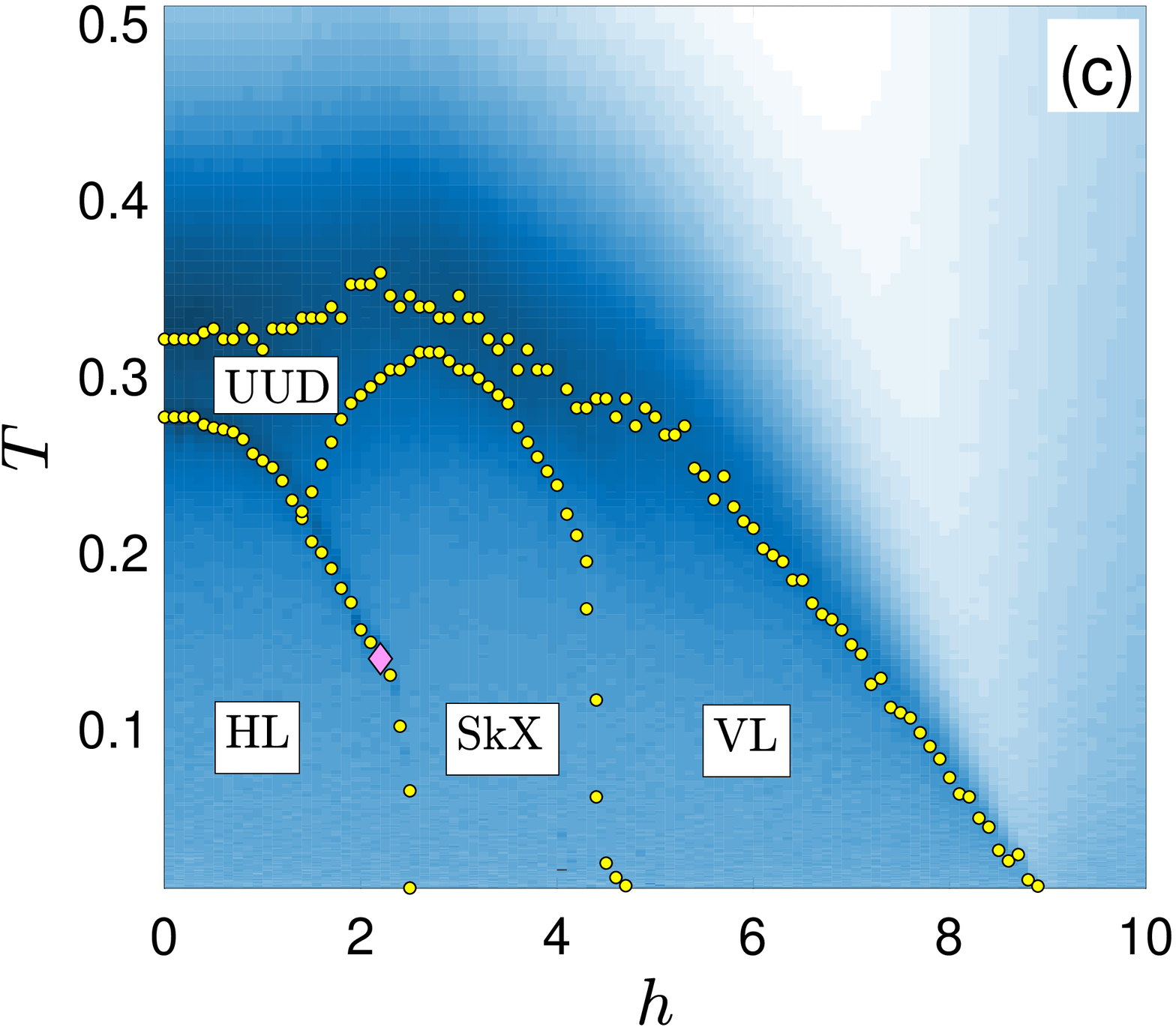}\label{fig:D020}}
\subfigure{\includegraphics[scale=0.32,clip]{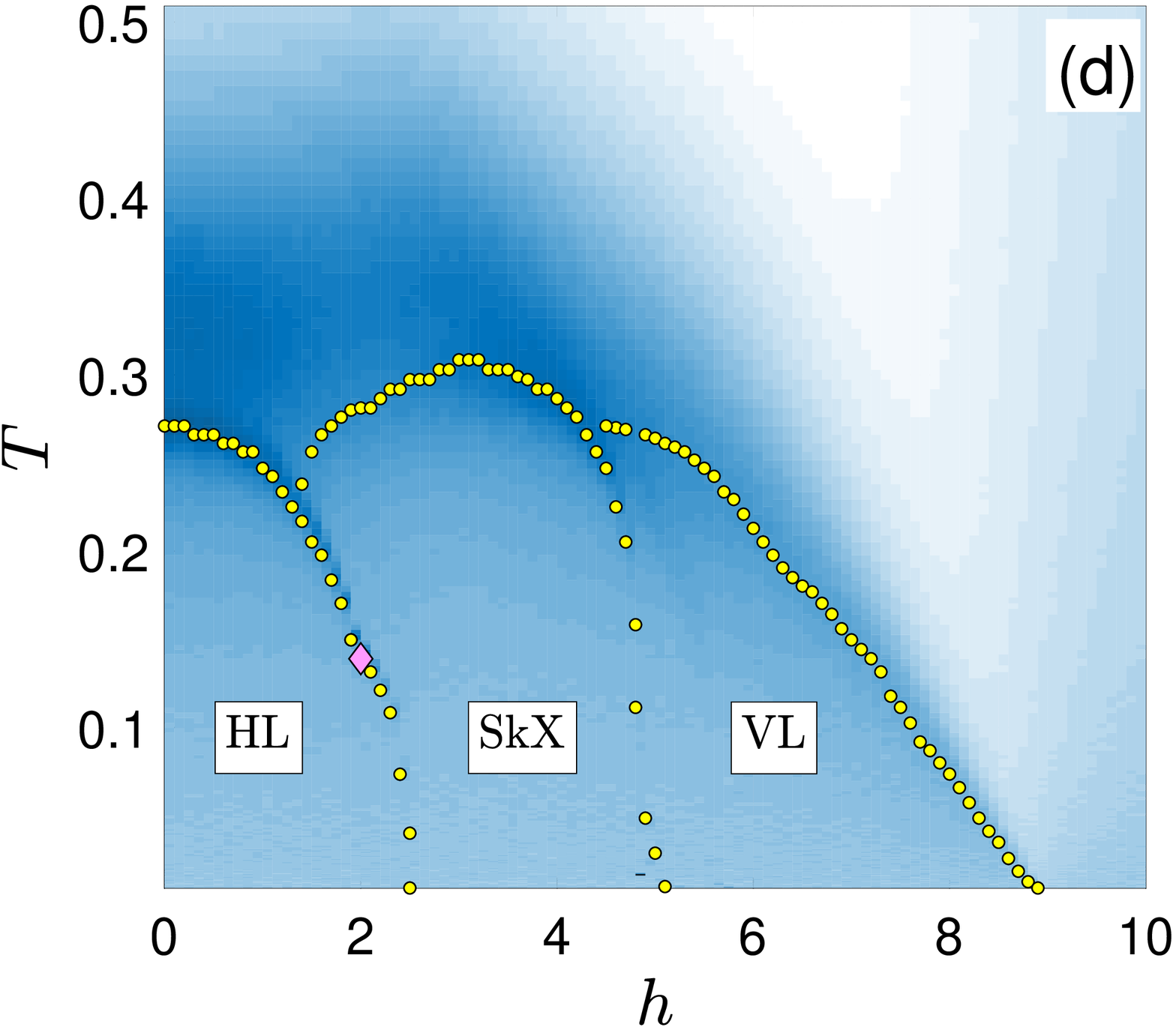}\label{fig:D030}}
\caption{Evolution of the phase diagram topology in $T-h$ plane for $L = 48$ and smaller $D$ values: (a) $D=0$, (b) $D=0.1$, (c) $D=0.2$, and (d) $D=0.3$.}
\label{fig:PD_small_D}
\end{figure}

\begin{figure}[t!]
\centering
\subfigure{\includegraphics[scale=0.32,clip]{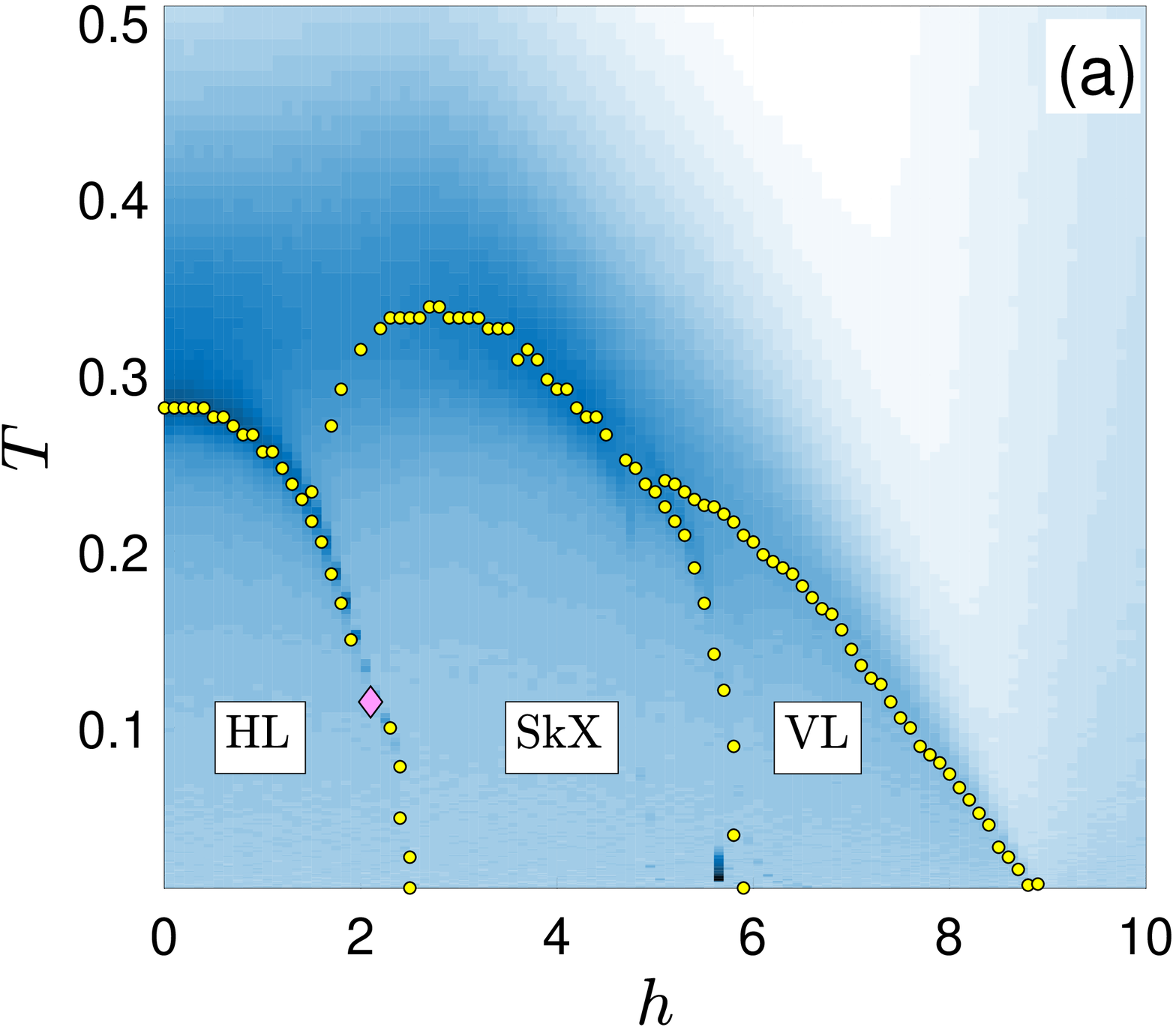}\label{fig:D040}}
\subfigure{\includegraphics[scale=0.32,clip]{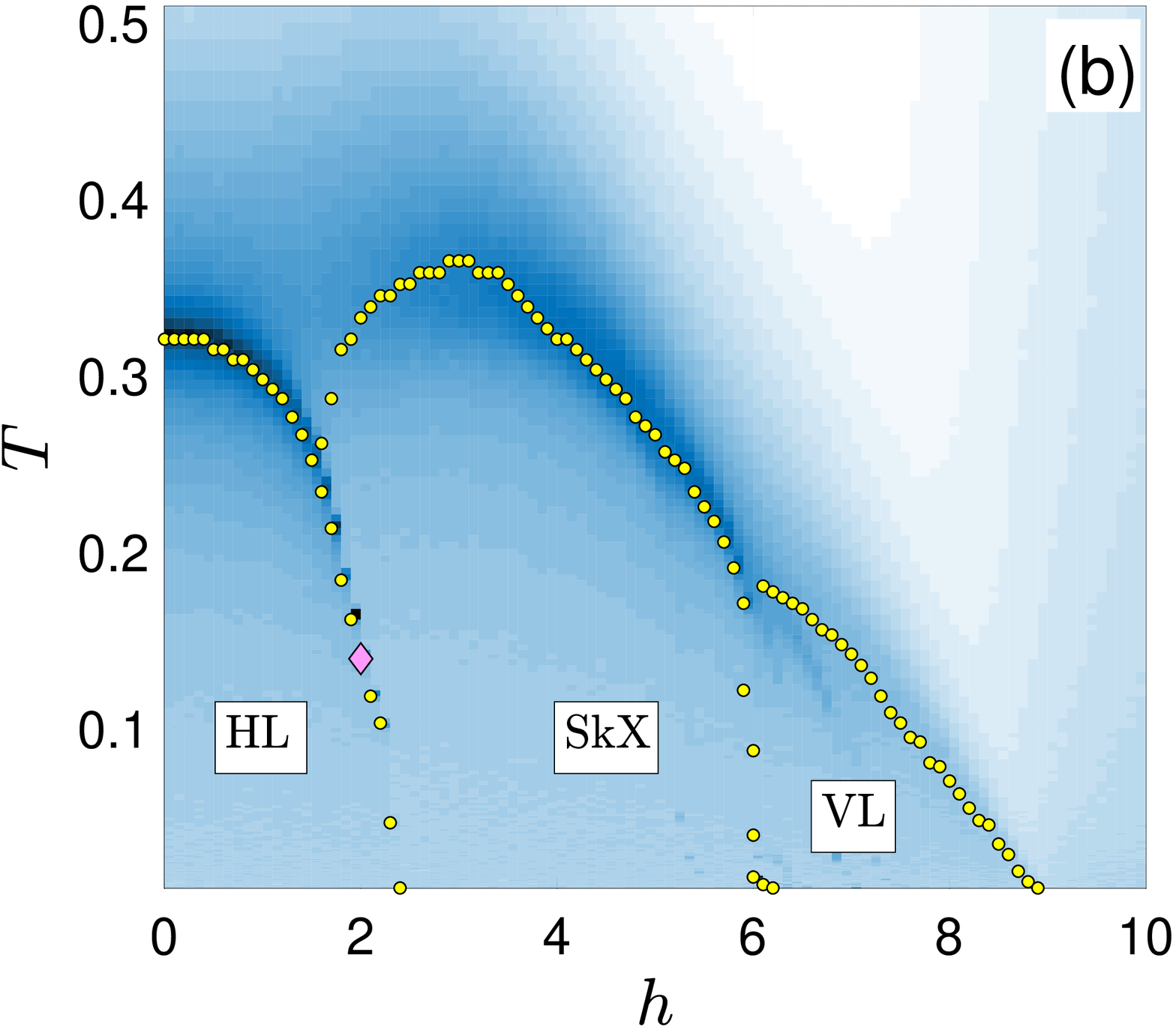}\label{fig:D060}}\\
\subfigure{\includegraphics[scale=0.32,clip]{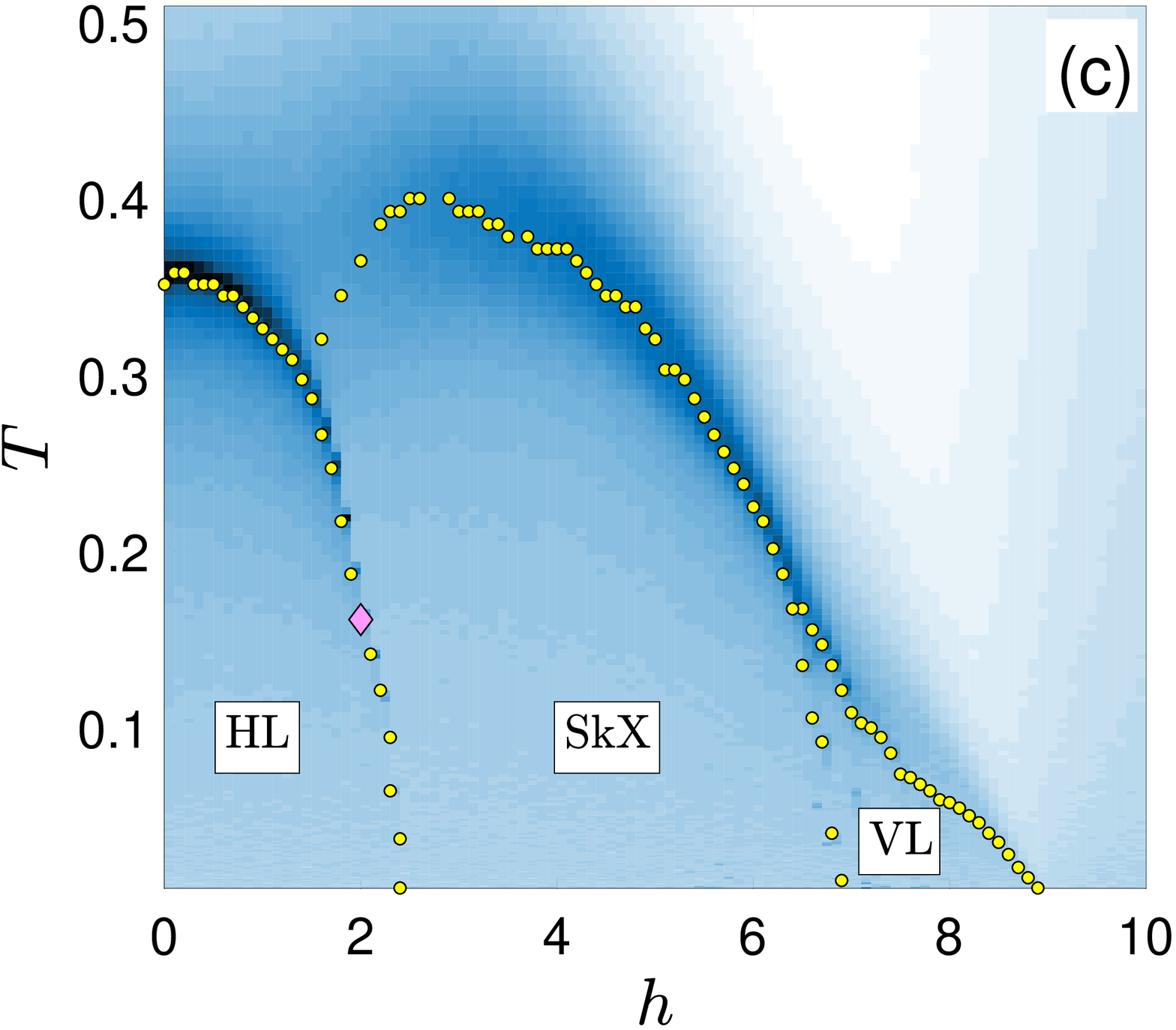}\label{fig:D080}}
\subfigure{\includegraphics[scale=0.32,clip]{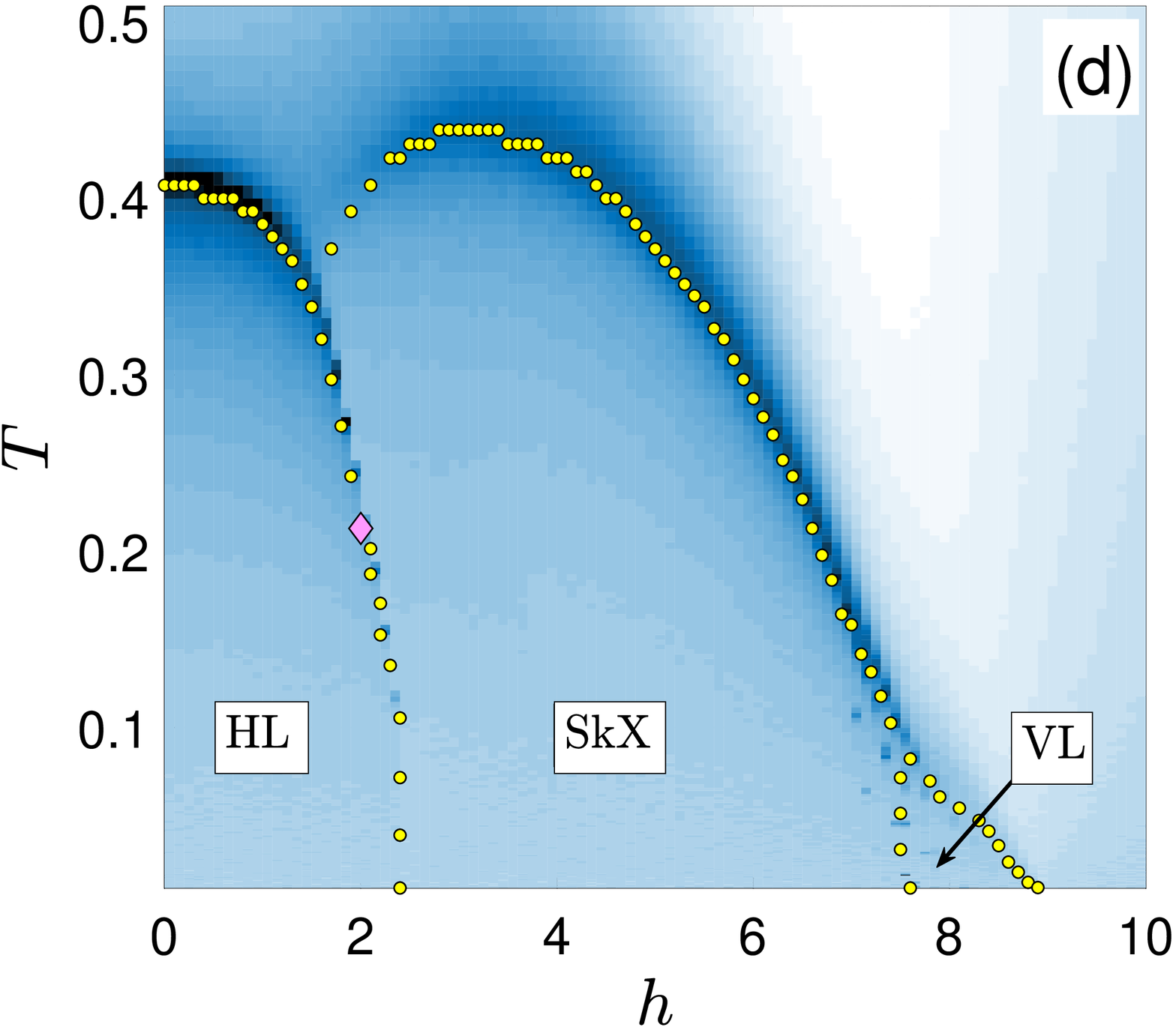}\label{fig:D100}}
\caption{Phase diagrams in $T-h$ plane for $L = 48$ and larger $D$ values: (a) $D=0.4$, (b) $D=0.6$, (c) $D=0.8$, and (d) $D=1.0$.}
\label{fig:PD_large_D}
\end{figure}

In order to construct phase diagrams in $T-h$ plane in the presence of the DMI with a varying intensity $D$, depending on circumstances in different regions of the parameter space, we combine the information obtained from the peaks of the calculated response functions, the magnetization and the SkX lattice order parameters, as well as spin snapshots. In Fig.~\ref{fig:PD_small_D} we present the evolution of the phase diagram topology in $T-h$ plane for smaller $D$ values, starting from $D=0$. We note that, strictly speaking, the presented diagrams are actually pseudo phase diagrams, obtained for a finite-size lattice corresponding to $L = 48$. Therefore, similarly to the $D=0$ case~\cite{seabra_2011}, caution should be exercised when extrapolating the results to the thermodynamic limit. The (pseudo)transition temperatures, denoted by the small yellow circles, are overlaid on the specific heat surfaces with darker shades corresponding to larger (peak) values. As one can see, inclusion of a relatively small DMI leads to the change of the phase diagram topology. The phase diagram for $D=0.1$, shown in Fig.~\ref{fig:D010}, features four ordered phases: the helical (HL) phase, the up-up-down (UUD), the V-like (VL), and the skyrmion lattice (SkX) phases. The SkX opens up at very low temperatures near $h \approx 3$, wedged between the HL and VL phases. With the increasing $D$ the area of the SkX phase grows and extends to higher temperatures, mainly at the cost of the UUD and VL phases. We note that the peaks of the response functions, related to the order-disorder transitions, become rather broad and scattered, which makes it difficult to determine the corresponding phase boundaries with higher precision. Therefore, for example in Fig.~\ref{fig:D020}, we expect the presence of the phase boundary between the UUD and VL phases but its location is very difficult to find. For $D=0.3$ the UUD phase has completely disappeared and there is a direct transition from the SkX to the P phase. The magenta diamond symbols on the HL-SkX phase boundaries denote approximate locations of the tricritical points, separating the first-order phase transitions at lower temperatures from the second-order ones at higher temperatures. We note that for $D=0.1$ (Fig.~\ref{fig:D010}) the whole HL-SkX phase boundary appears to be of the first order.

\begin{figure}[t!]
\centering
\subfigure{\includegraphics[scale=0.32,clip]{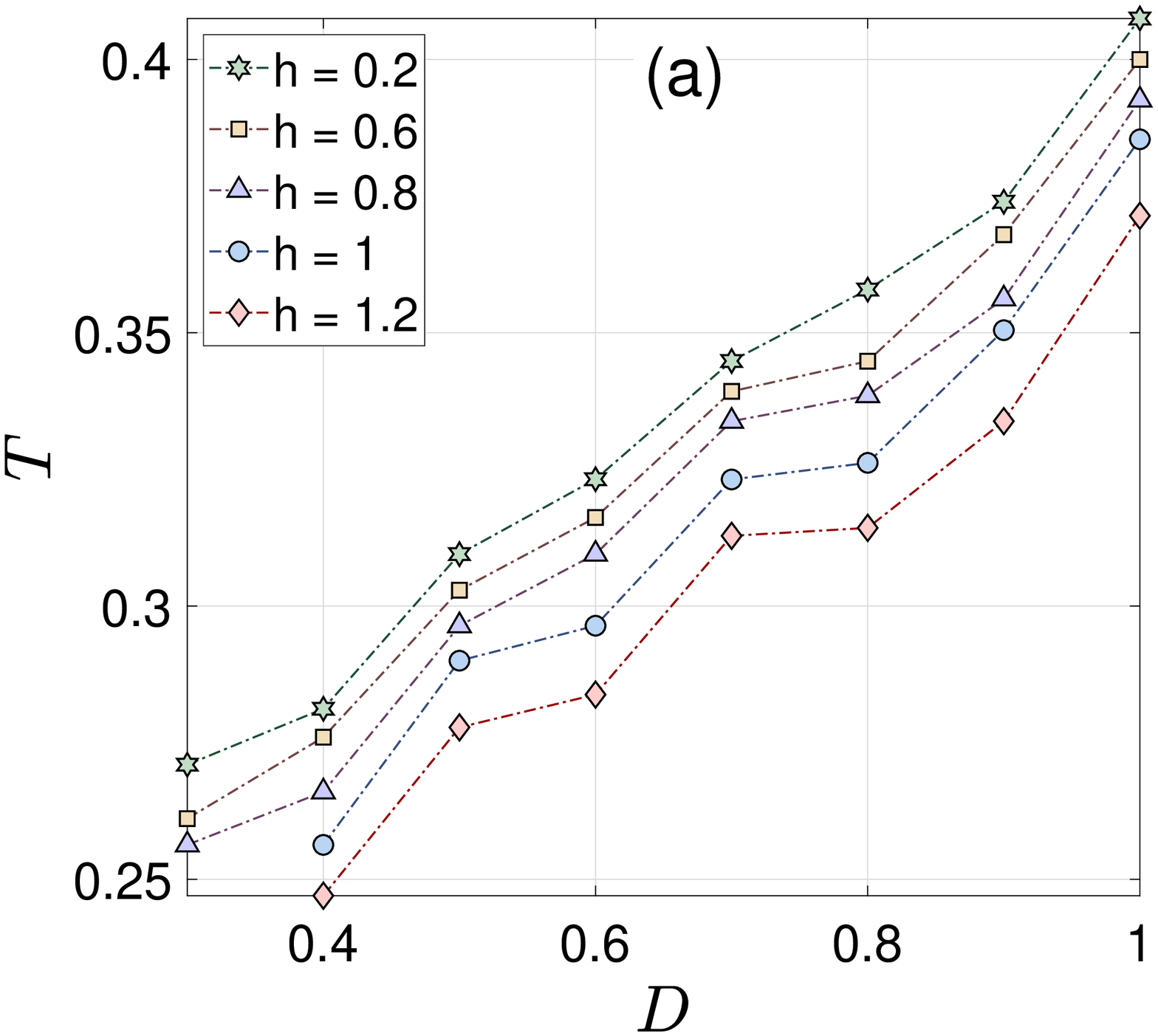}\label{fig:para_to_hel}}
\subfigure{\includegraphics[scale=0.32,clip]{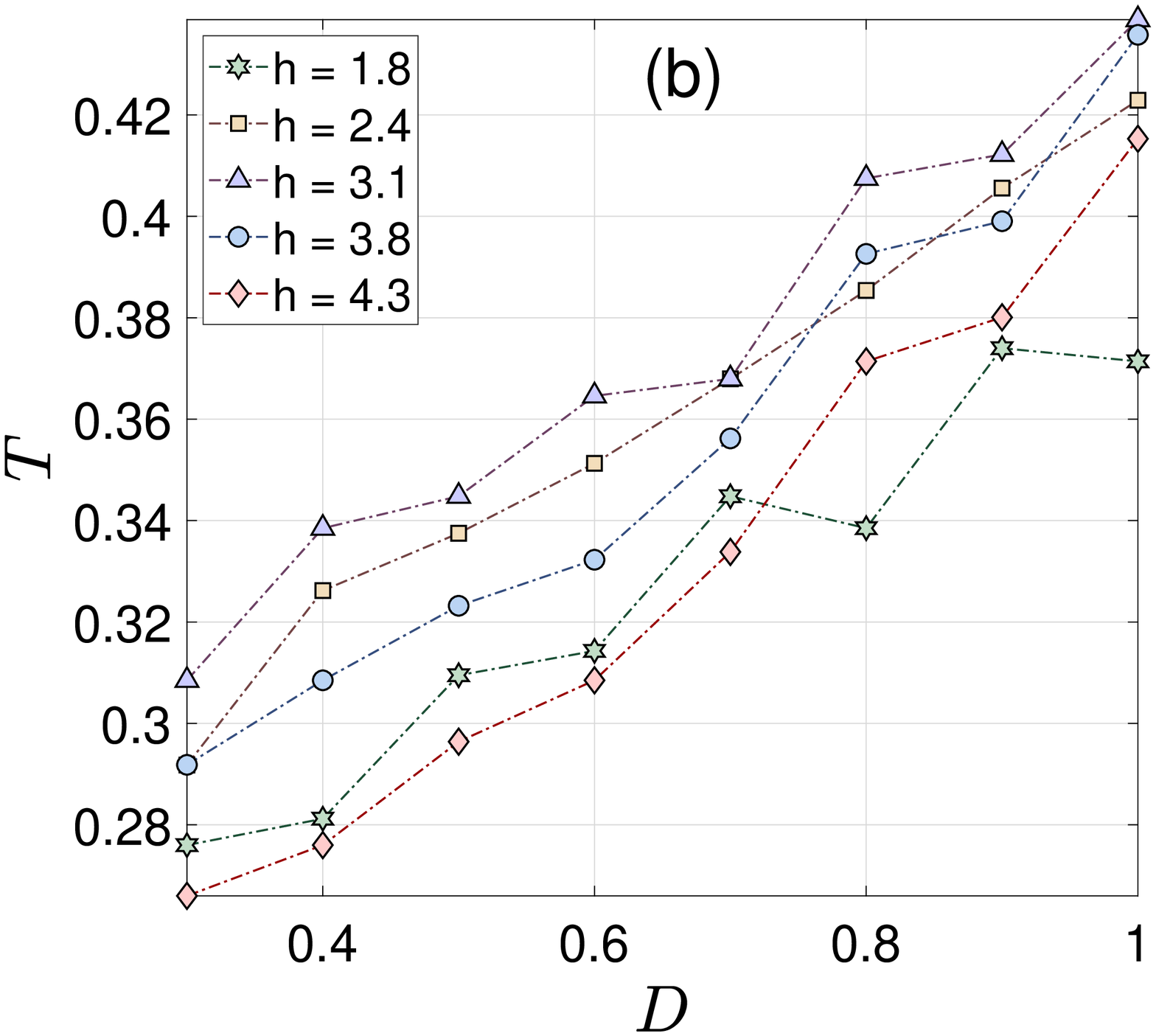}\label{fig:para_to_sk}}\\
\subfigure{\includegraphics[scale=0.32,clip]{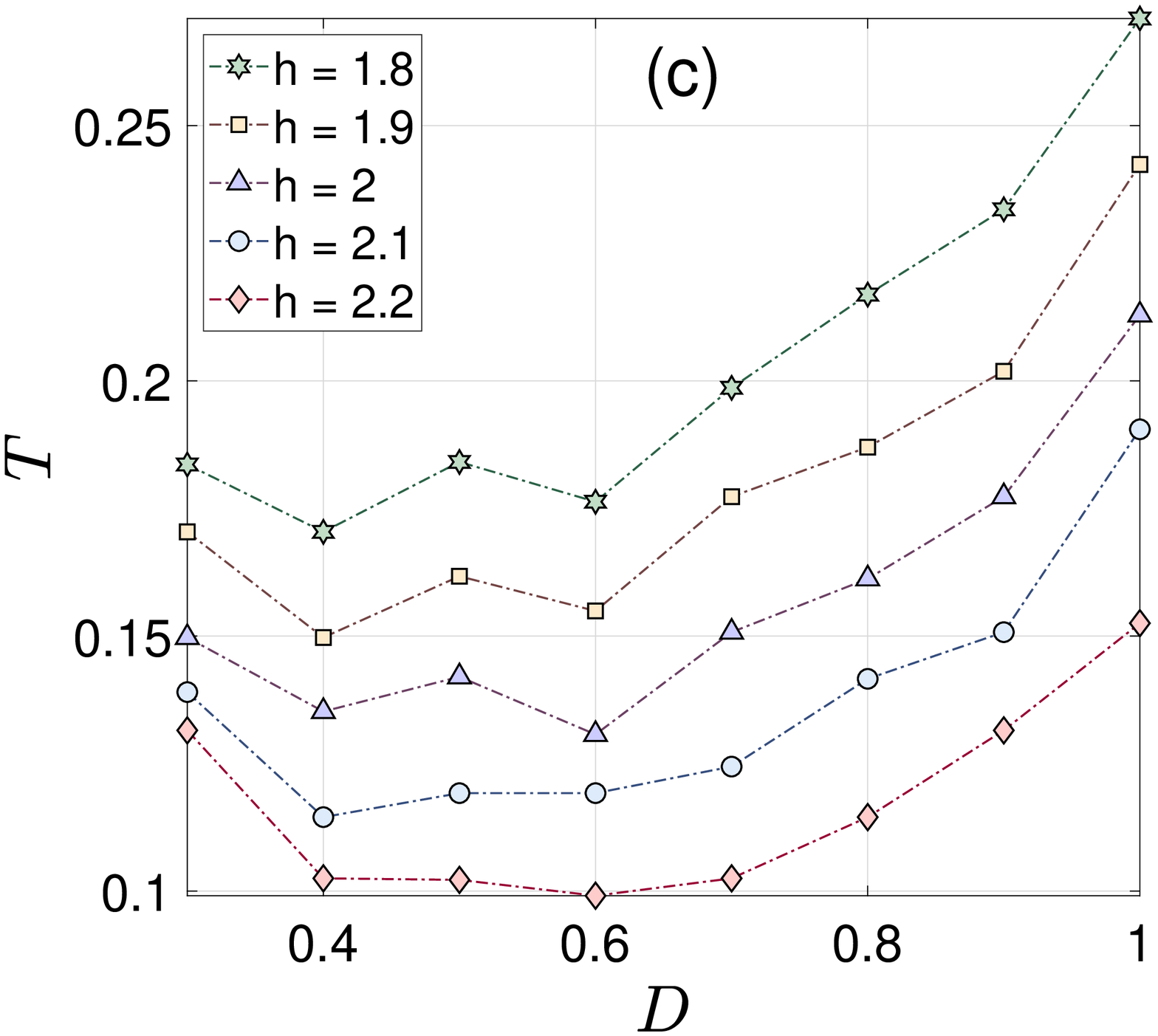}\label{fig:hel_to_sk}}
\subfigure{\includegraphics[scale=0.32,clip]{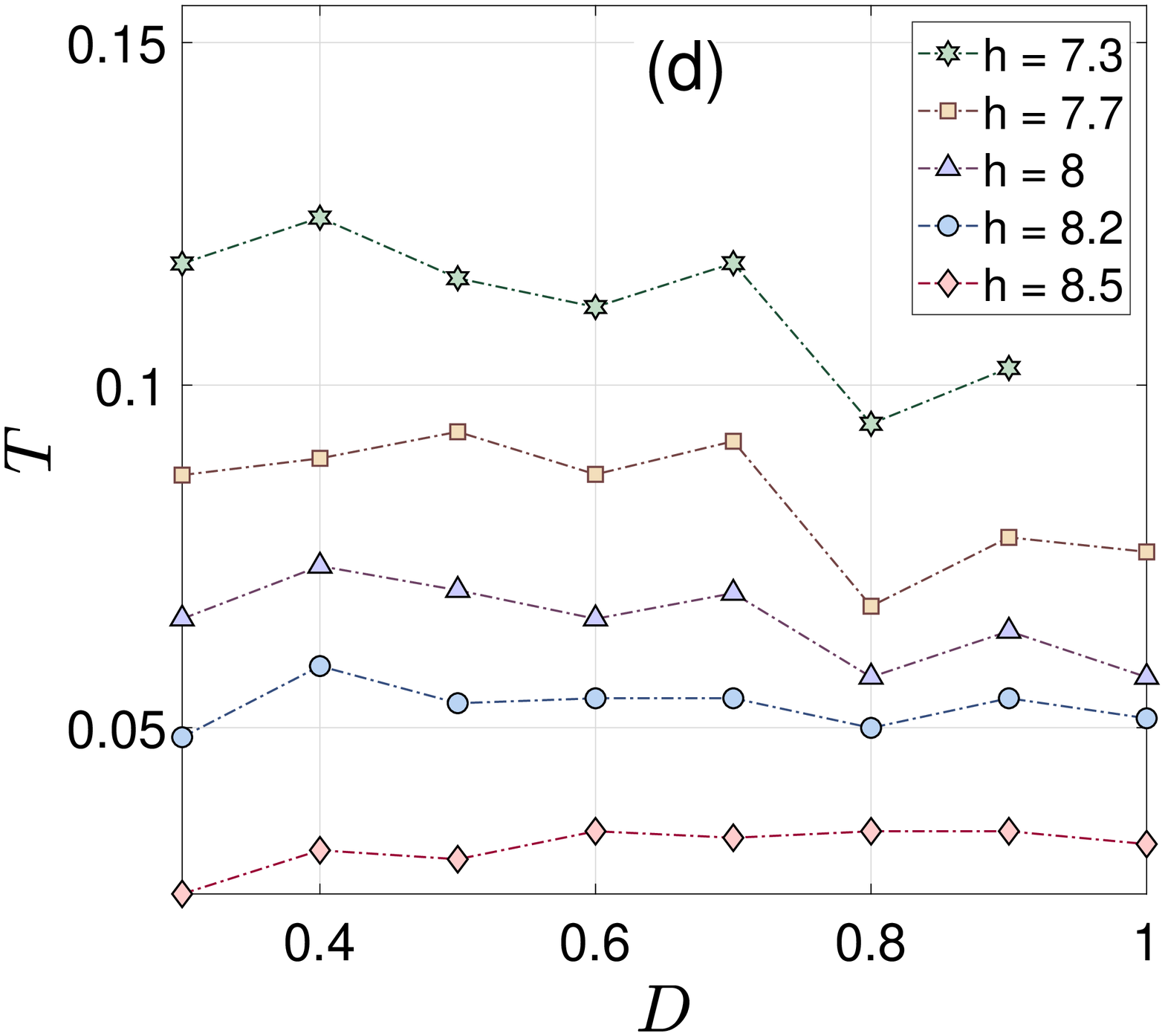}\label{fig:para_to_SU}}
\caption{Transition temperatures as functions of the DMI parameter $D$ for selected field values at (a) P-HL, (b) P-SkX, (c) HL-SkX, and (d) P-VL phase transitions.}
\label{fig:PD_D_dep}
\end{figure}

Figure~\ref{fig:PD_large_D} shows that for larger values of $D$ (up to at least $D=1$), the topology of the phase diagrams does not change. Nevertheless, there are quantitative changes, which include shrinking of the VL phase and shifting it to larger field values. Dependencies of the transition temperatures between different phases on the DMI intensity are presented in Fig.~\ref{fig:PD_D_dep}, for several corresponding field values. Apparently, the increasing DMI stabilizes both the HL and the SkX phases by increasing the transition temperatures from these phases to the disordered P phase (Figs.~\ref{fig:para_to_hel} and~\ref{fig:para_to_sk}). Similar tendency can be observed also for the HL-SkX transition, however only starting from intermediate values of $D$. For smaller $D$ its increase makes the HL-SkX transition temperatures decrease, thus making the whole transition boundary asymmetrically U-shaped (Fig.~\ref{fig:hel_to_sk}). On the other hand, the effect of the increasing $D$ on the P-VL transition temperature is not as apparent. Nevertheless, some decreasing tendency for smaller fields, which becomes less and less pronounced and even slightly increasing for extremely large fields approaching the saturation value $h=9$, can be observed (Fig.~\ref{fig:para_to_SU}). 

\section{Summary and discussion}

We have considered a frustrated Heisenberg antiferromagnet in an external magnetic field and studied the formation and evolution of the antiferromagnetic skyrmion crystal (SkX) with the increasing Dzyaloshinskii-Moriya interaction (DMI) intensity. By studying the model in a wide parameter space we obtained a rather comprehensive picture of its critical behavior. It was found that already relatively small DMI intensity of $D_t \approx 0.02$ can lead to the appearance of the SkX phase in the vicinity of the meeting point of the remaining three ordered phases, present also for $D < D_t$: the helical (HL), the coplanar up-up-down (UUD) and the V-like (VL). 

\begin{figure}[t!]
\centering
\subfigure{\includegraphics[scale=0.32,clip]{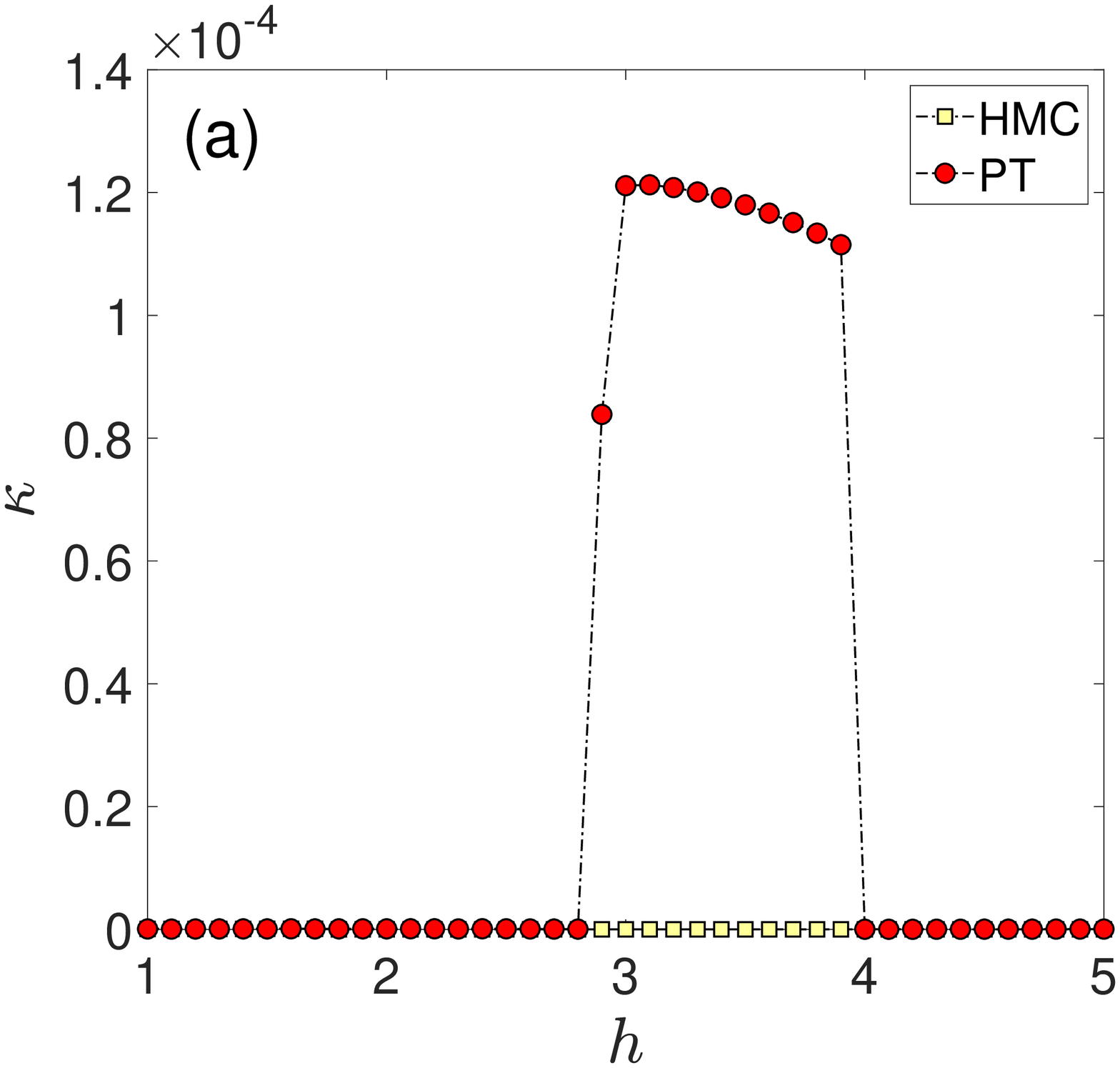}\label{fig:D008_xiL}}
\subfigure{\includegraphics[scale=0.32,clip]{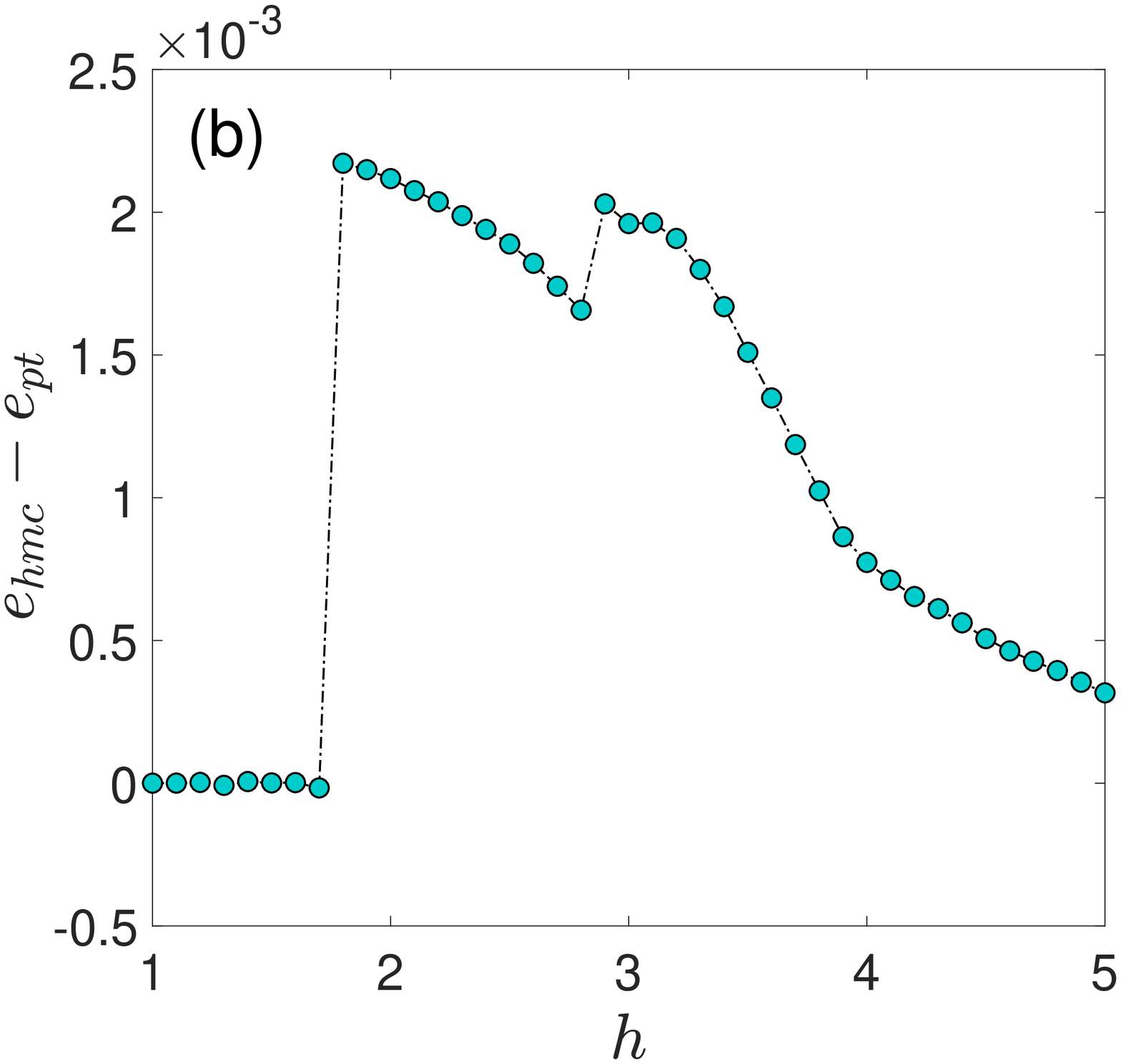}\label{fig:D008_Edif}}
\subfigure{\includegraphics[scale=0.32,clip]{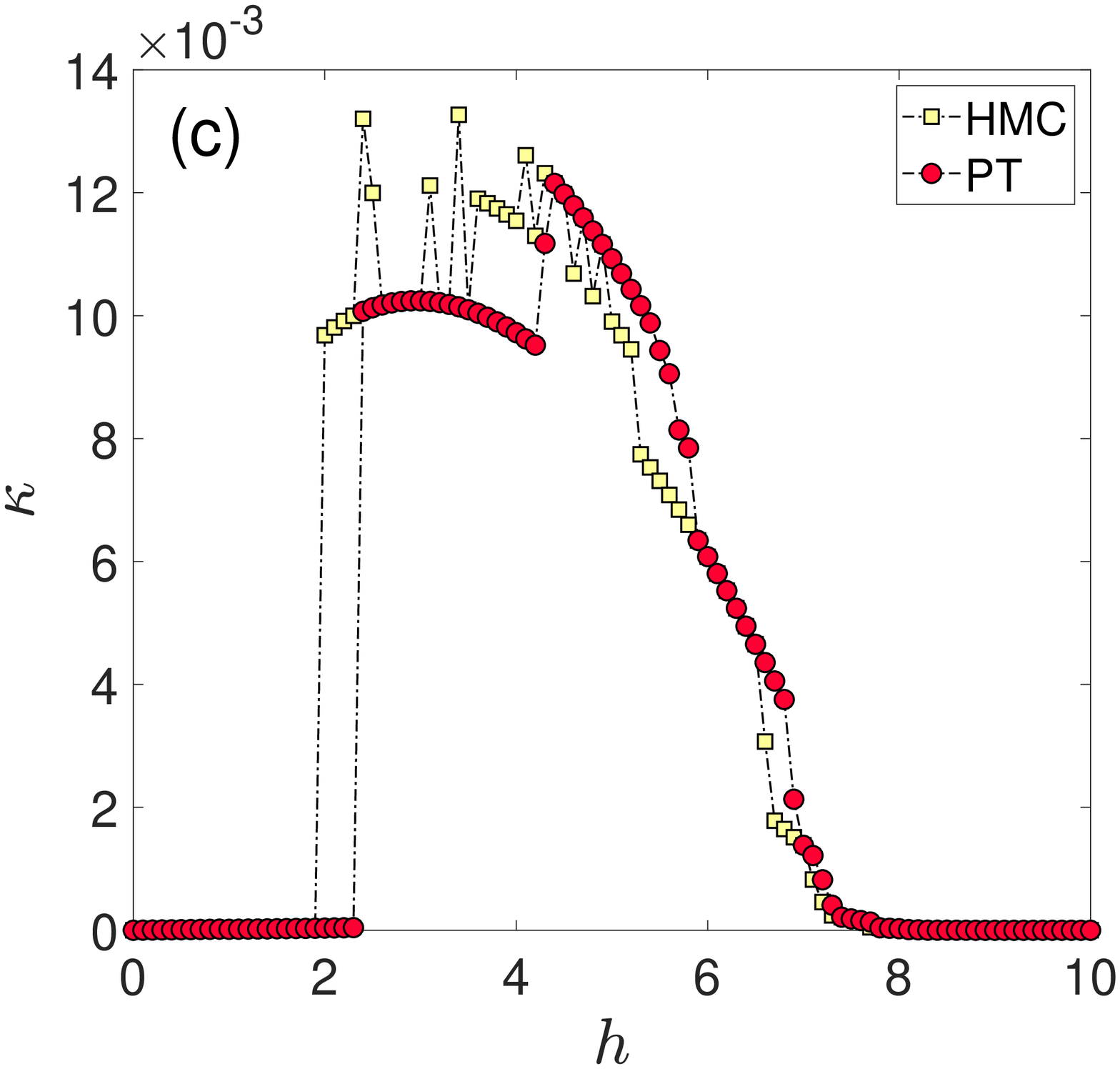}\label{fig:D080_xiL}}
\subfigure{\includegraphics[scale=0.32,clip]{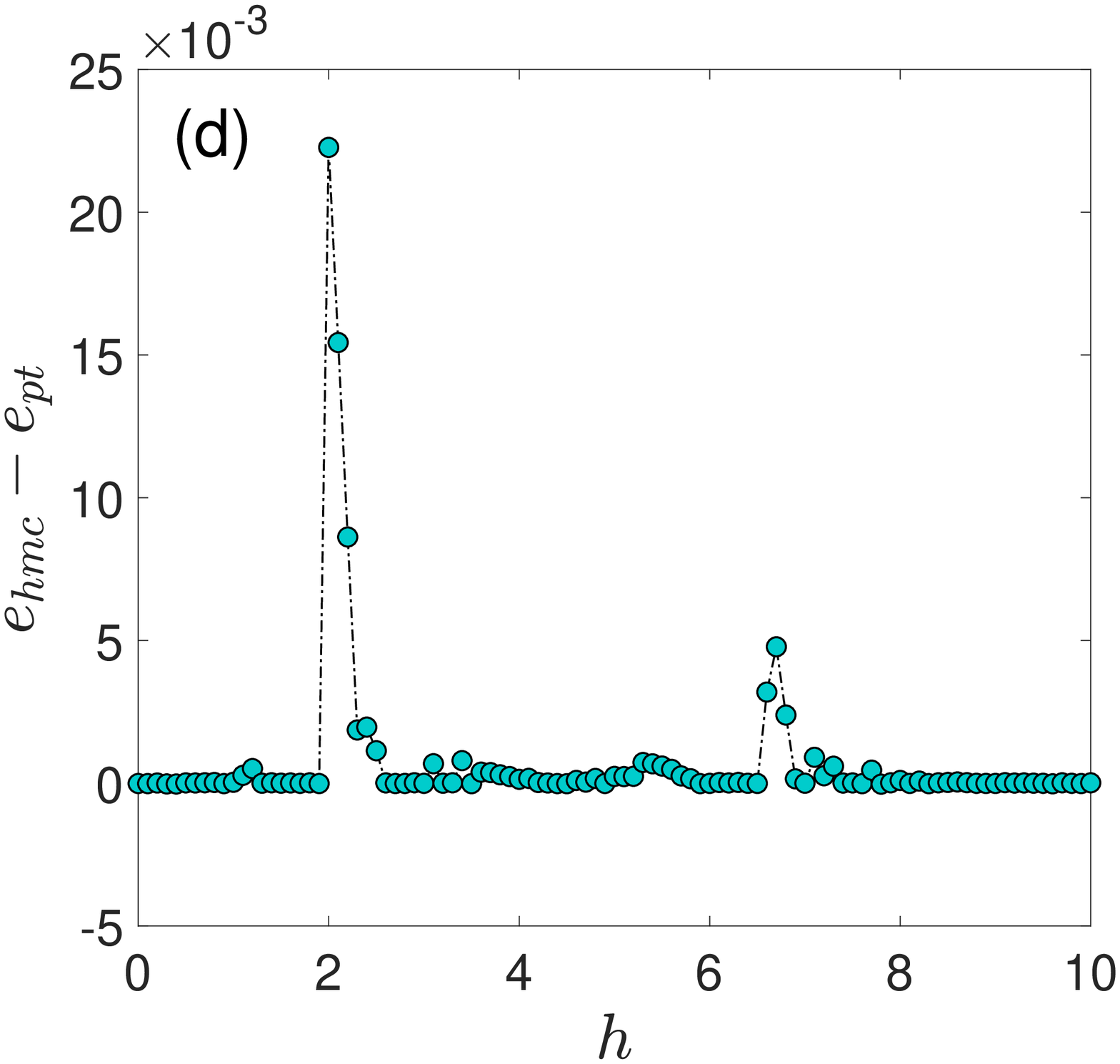}\label{fig:D080_Edif}}\\
\caption{Comparison of the performance of the HMC and PT methods at $T=0.009$ in terms of reliability of (a,b) the SkX phase detection at very low $D$ and (c,d) determination of the first-order HL-SkX phase boundary. Panels (a) and (c) show the field dependencies of the chiralities, for $D = 0.08$ and $D = 0.8$, respectively, and (b) and (d) the energy differences between the HMC and PT solutions, where $e=\langle \mathcal{H} \rangle/N$.}
\label{fig:SMC_vs_PT}
\end{figure}

For $D > D_t$ the phase diagram topology was found to evolve up to $D \approx 0.3$, beyond which (up to $D = 1$) only quantitative changes could be observed. The HL-SkX phase boundary was confirmed to include first-order phase transitions at low temperatures, which for $D \geq 0.2$ with increasing temperature changed to the second-order ones at trictitical points. We also studied the effect of the field and DMI magnitudes on the skyrmion size. In line with some previous studies on the present frustrated antiferromagnetic system~\cite{rosales,fr} but in contrast to the results obtained for some ferromagnetic thin film systems in the presence of an interfacial DMI~\cite{fatt18,fatt20}, we observed that the increasing field tends to increase the skyrmion size, albeit not necessarily for larger $D$ (see Fig.~\ref{fig:D080_xiL}). On the other hand, we found that the DMI has just the opposite effect, i.e., increasing its magnitude makes skyrmions smaller down to the smallest size, reached at $D \approx 0.94$, when skyrmions consist of only one central spin pointing opposite to the field direction and six surrounding spins lying in the plane. 

We have employed the parallel tempering (PT) method, which enabled us to obtain more reliable and precise results than those previously obtained by applying an approximate low-energy effective theory~\cite{osorio_17} and the standard Monte Carlo (MC) simulation~\cite{mohylna_20}. One example is the presence of the SkX phase for relatively small DMI, $0.04 < D < 0.2$, which could not be detected by the above methods but it was revealed by PT. In Fig.~\ref{fig:D008_xiL} we demonstrate that, for $D=0.08$, $T= 0.009$ and $L = 90$, the field dependence of the chirality obtained by the standard HMC is constantly zero, indicating the absence of the SkX phase\footnote{Closer analysis reveals the UUD phase, within $1.8 \leq h \leq 2.8$, and the VL phase, for $h > 2.8$.}, while the one obtained by PT indicates the presence of the SkX phase within $3 \leq h < 4$. Fig.~\ref{fig:D008_Edif} shows, that PT provides a more stable solution while HMC got trapped in a local minimum, which is very close to the global one since the difference between the energy levels is very small.

%%%%%%%%%%%%MT PT DIF

Another example of the superiority of PT over the standard HMC is presented in Fig.~\ref{fig:D080_xiL} and it is mainly related to the location of the first-order phase transition. It shows a similar dependence but now for $D=0.8$. In this case the standard HMC managed to capture the SkX phase but apparently overestimated the field interval of its existence. Based on HMC the interval $2 \leq h \leq 2.3$ corresponds to the SkX phase but PT shows that this solution is only metastable, separated from the stable one by a relatively large energy barrier (see Fig~\ref{fig:D080_Edif}), and the stable solution corresponds to the HL phase with zero chirality. In HMC we can also observe within the SkX phase switching between different solutions, corresponding to the states with larger number of smaller size skyrmions and smaller number of larger size skyrmions, which are separated by very small energy barriers. Those can be reliably tunneled through using PT, finding the lowest-energy stable state. Fig.~\ref{fig:D080_xiL} also shows one peculiarity in the PT solution. As discussed above, the increasing field typically increases the size of skyrmions, which translates to the decease of the skyrmion number and the chirality in the vicinity of the SkX-VL transition boundary. However, here we can witness also the opposite phenomenon, i.e., the increase of these quantities, well within the SkX phase (at $h \approx 4.3$). In our simulations we observed that it occurs only at larger DMI intensity ($D>0.5$), as a result of a close proximity of the energy levels of different (apparently for $D=0.8$ at least three) solutions.

\section*{Acknowledgment}
This work was supported by the Scientific Grant Agency of Ministry of Education of Slovak Republic (Grant No. 1/0531/19) and the Slovak Research and Development Agency (Contract No. APVV-16-0186). Part of computations was held on the basis of the HybriLIT heterogeneous computing platform (LIT, JINR)~\cite{hybrilit}. 
%M. M. expresses the deepest gratitude to J\'{a}n Bu\v{s}a Jr. for patient and essential guidance on parallel programming.

%\bibliographystyle{elsarticle-num}
%\bibliography{pd}

\end{document}